\documentclass{JHEP3}
\usepackage{times}
\usepackage{graphics}
\usepackage{graphicx}
\usepackage{epsf}
\usepackage{amsmath}

\usepackage{graphicx}
\usepackage{rotating}

\numberwithin{equation}{section}
\numberwithin{figure}{section}
\numberwithin{table}{section}

\usepackage{epsfig}
\usepackage{rotating}

\def \eps {\epsilon}
\def \veps  {\varepsilon}

\newcommand{\be}{\begin{equation}}
\newcommand{\ee}{\end{equation}}
\newcommand{\bea}{\begin{eqnarray}}
\newcommand{\eea}{\end{eqnarray}}
\newcommand{\non}{\nonumber}
\newcommand{\nl}{\nonumber \\}

\newcommand{\nn}{\nonumber}

\title{
\vspace*{-2cm}
{\small\texttt{DESY 08-174}}\\[-4mm]
{\small\texttt{BI-TP 2008/39}}\\[-4mm]
{\small\texttt{SFB-CPP-08-97}}\\[-4mm]
{\small\texttt{HEPTOOLS 08-046}}
\\[5mm]
A complete reduction of one-loop tensor \\5- and 6-point integrals}

\author{
Th. Diakonidis~${}^{a}$, J. Fleischer~${}^{a,b}$,
J. Gluza~${}^{c}$, K. Kajda~${}^{c}$, T. Riemann~${}^{a}$,
J. B. Tausk~${}^{a}$ \\

$^{a}$~Deutsches Elektronen-Synchrotron, DESY, Platanenallee
  6, 15738 Zeuthen, Germany \\
$^{b}$~Fakult\"at f\"ur Physik, Universit\"at Bielefeld, Universit\"atsstr. 25,  33615
Bielefeld, Germany \\
$^{c}$~Institute of Physics, University of Silesia,
Uniwersytecka 4, 40-007 Katowice, Poland}

\abstract{
We perform a complete analytical reduction of general one-loop Feynman
integrals with five and six external legs for tensors up to rank $R=3$
and $4$, respectively. An elegant formalism with extensive use of signed
minors is developed for the cancellation of inverse Gram determinants.
The 6-point tensor functions of rank $R$ are expressed in terms of
5-point tensor functions  of rank $R-1$, and the latter are reduced to
scalar four-, three-, and two-point functions. The resulting compact
formulae allow both for a study of analytical properties and for efficient
numerical programming. They are implemented in Fortran and Mathematica.
}

\preprint{}

\begin{document}
\allowdisplaybreaks





\clearpage

\section{\label{Intro} Introduction}
At the proton-proton collider LHC and the planned $e^+e^-$ collider ILC,
a large number of particles per event may be produced.
The hope is to discover one or several Higgs bosons or supersymmetric
particles, which are typically expected to be quite heavy.
The interest is also directed to the study of known massive particles
like the $W$ and $Z$ bosons or the top quark.
Since the production rates are large, a proper description of the
cross-sections will typically include one-loop corrections to $n$-particle
reactions, where some of the final state particles may be massive.

The Feynman integrals for reactions with up to four external particles
have been systematically studied and evaluated in numerous studies.
We just want to mention here the seminal papers \cite{'tHooft:1979xw} and
\cite{Passarino:1979jh}
and the Fortran packages \texttt{FF} \cite{vanOldenborgh:1990yc} and
\texttt{LoopTools} \cite{Hahn:1998yk}, which represent the state of the
art until now.
The treatment of Feynman integrals with a higher multiplicity than
four becomes quite involved if questions of efficiency and stability
become vital, as it happens with the calculational problems related to
high-dimensional phase space integrals over sums of thousands of Feynman
diagrams with internal loops.

In this article, we will concentrate on the evaluation of massive one-loop
Feynman integrals with $n$ external legs and some tensor structure,
\begin{eqnarray}\label{definition}
 I_n^{\mu_1\cdots\mu_R} &=&  ~~\int \frac{d^d k}{i\pi^{d/2}}~~\frac{\prod_{r=1}^{R} k^{\mu_r}}{\prod_{j=1}^{n}c_j^{\nu_j}},
\end{eqnarray}
where the denominators $c_j$ have \emph{indices} $\nu_j$ and \emph{chords}
$q_j$,
\begin{eqnarray}\label{propagators}
c_j &=& (k-q_j)^2-m_j^2 +i\veps .
\end{eqnarray}
We will study in the following the cases $n=5$ with $R\leq 3$ and $n=6$
with $R\leq 4$, and we will conventionally assume $q_n=0$.
The space-time dimension is $d=4-2\eps$.

There are several strategies one might follow.
One is the reduction of higher-point tensor
integrals to tensor integrals with less external lines and/or lower tensor rank
\cite{Denner:2002ii,Denner:2005nn,Binoth:1999sp,Binoth:2005ff};
a second approach is essentially  numerical \cite{Ferroglia:2002mz,Kurihara:2005ja} or semi-numerical  \cite{Ossola:2006us,Ossola:2007ax,Ellis:2005zh}.
A third one rests on the unitarity cut method \cite{Bern:1996je,Dixon:1996wi,Bern:2007dw,Giele:2008bc}. 
In this case, a one-loop amplitude is evaluated as a whole,  by using Cutkosky rules,
instead of computing loop integrals from each of the Feynman diagrams.
It is impossible to give here a comprehensive survey
of recent activities, and we would like to refer to
e.g. \cite{Weinzierl:2007vk,Bern:2008ef,LL:2008xx,Binoth:2008zz} for
recent overviews on the subject.

Here, we will advocate yet another approach and reduce the tensor
integrals algebraically to sums over a small set of scalar two-, three-
and four-point functions, which we assume to be known.  Whether such
a complete reduction is competitive with the other approaches might
be disputed. Evidently, this depends on the specific problem under
investigation. For a study of gauge invariance and of the ultraviolet
(UV) and infrared (IR) singularity structure of a set of Feynman diagrams,
it is evident that a complete reduction  is advantageous, and it may
also be quite useful for a tuned, analytical study of certain regions
of potential numerical instabilities.

We have chosen a strictly algebraic approach and will rely heavily on
the algebra of signed minors which was worked out in detail by Melrose
in \cite{Melrose:1965kb}.  One of the basic observations of Melrose was
that in four dimensions all the scalar integrals can be reduced to scalar
4-point functions and simpler ones. 
In \cite{Davydychev:1991va}, a representation of arbitrary one-loop tensor
integrals in terms of scalar integrals was derived.  The  representation
includes, however, scalar integrals with higher indices ${\nu}_j$ and
higher space-time dimensions $d+2l$. The subsequent reduction to scalar
integrals with only the original indices and the generic  space-time
dimension $d$ is possible with the use of integration-by-parts identities
\cite{Chetyrkin:1981qh} and generalizations of them with dimensional
shifts. The latter have been derived in \cite{Tarasov:1996br},
and a systematical application to one-loop integrals may be found in
\cite{Fleischer:1999hq}.%
\footnote{We will extensively quote from article \cite{Fleischer:1999hq},
so we introduce here the notation (I.num) for a reference to equation
(num) there.}
Basically, the reduction problem has been solved this way for $n$-point
functions.
There was one attempt to use the Davydychev-Tarasov reduction for the
description of one-loop contributions to the process
$e^+e^-\to H\nu{\bar \nu}$  \cite{Jegerlehner:2002es}, and 
the numerical problems due to the five-point functions were discussed
in some detail.
To a large extent they root in the appearance of inverse powers of Gram
determinants.
This feature of the Davydychev-Tarasov reduction was identified as
disadvantageous soon after its derivation, e.g. in \cite{Campbell:1996zw},
where a strategy for avoiding these problems was developed.
Besides the problem of inverse powers of the Gram determinant of
the corresponding Feynman diagram, there are additional kinematical
singularities related to sub-diagrams.
This will not be discussed here; we refer to e.g.
\cite{Campbell:1996zw,Devaraj:1998es,%
Denner:2002ii,Denner:2005nn,Binoth:1999sp,Binoth:2005ff,Ellis:2005zh,%
Giele:2008bc} and references therein.

In this article, we investigate the reduction of tensor integrals with five and six external legs which are of immediate importance in applications at the LHC.
In Section \ref{secFJT} we represent tensor integrals by scalar integrals
in shifted space-time dimensions with shifted indices.
Section \ref{rank1234} and Section \ref{tens6} contain our main result.
In Section \ref{rank1234} we 
go one step further in  the reduction of five-point tensors  compared to \cite{Fleischer:1999hq} and
demonstrate how to cancel all inverse powers of the Gram determinant appearing in the  Davydychev-Tarasov reduction.
Earlier results for tensors of rank two may be found in \cite{Fleischer:2007ph}.
Section \ref{tens6} contains the reduction of 
tensorial six-point functions to tensorial 5-point functions. 
The
corresponding Gram determinant is identically zero
\cite{Fleischer:1999hq,Denner:2005nn,Binoth:2005ff}, and the reduction becomes quite compact.
Some numerical results and a short discussion are given in
Section \ref{Numerical}.
The numerics is obtained with two independent implementations, one made
in Mathematica, and another one in Fortran.
The Mathematica program \texttt{hexagon.m} with the reduction formulae
is made publicly available
\cite{hexagon:2008}, see also \cite{Diakonidis:2008dt} for a short description.
For numerical applications, one has to link the package with a program for the
evaluation of scalar one- to four-point functions, e.g. with
\texttt{LoopTools} \cite{Hahn:1998yk,Hahn:2006qw,vanOldenborgh:1990yc},
\texttt{CutTools} \cite{vanHameren:2005ed,Ossola:2007ax},
\texttt{QCDLoop} \cite{Ellis:2007qk}.
Appendices are devoted to some known, but necessary details on 
Gram determinants and the algebra of signed minors and to a short
summary about the reduction of dimensionally shifted four- and five-point
integrals.

\section{\label{secFJT} Representing tensor integrals by scalar integrals in shifted space-time dimensions}
At first we give the reduction of tensor integrals to a set of scalar integrals for arbitrary $n$-point functions. 
Following \cite{Davydychev:1991va,Fleischer:1999hq},
assuming here the indices of propagators to be equal to one,
$\nu_r=1$, one has:
\begin{eqnarray}
\label{scalarred}
  I_n^{\mu} & =&  \int ^{d} k^{\mu} \prod_{r=1}^{n} \, {c_r^{-1}} 
\nl& = &
-~\sum_{i=1}^{n-1} \, q_i^{\mu} \, I_{n,i}^{[d+]} ,
\\
\label{vectorred}
 I_{n}^{\mu\, \nu}
&=& 
\int ^{d} k^{\mu} \, k^{\nu} \, \prod_{r=1}^{n} \, {c_r^{-1}}
 \nl
 &=&  \sum_{i,j=1}^{n-1} \, q_i^{\mu}\, q_j^{\nu} \, n_{ij} \,  \, I_{n,ij}^{[d+]^2} -\frac{1}{2}
   \, g^{\mu \nu}  \, I_{n}^{[d+]} \, ,
 \\
\label{tensorred}
I_{n}^{\mu\, \nu\, \lambda}
&=& \int ^{d} k^{\mu} \, k^{\nu} \,  k^{\lambda} \, \prod_{r=1}^{n} \, {c_r^{-1}}
\nl 
 &=& -~ \sum_{i,j,k=1}^{n-1} \, q_i^{\mu}\, q_j^{\nu}\, q_k^{\lambda} \,  n_{ijk} \,  \, I_{n,ijk}^{[d+]^3}  
+
\frac{1}{2} \sum_{i=1}^{n-1} (\, g^{\mu \nu}  \, q_i^{\lambda} \,+ g^{\mu \lambda}  \, q_i^{\nu} \,+
    \, g^{\nu \lambda}  \, q_i^{\mu} \, )I_{n,i}^{[d+]^2} \, ,
\label{intone}
\\
I_{n}^{\mu \nu \lambda \rho} 
&=&
\int ^{d} k^{\mu}  k^{\nu}   k^{\lambda}  k^{\rho}  \prod_{r=1}^{n}  {c_r^{-1}}
\nl  
&=&  \sum_{i,j,k,l=1}^{n-1} \, q_i^{\mu}\, q_j^{\nu}\, q_k^{\lambda}
 \, q_l^{\rho}\,  n_{ijkl} \,  \, I_{n,ijkl}^{[d+]^4}  
\nn\\
&&    -~\frac{1}{2} \sum_{i,j=1}^{n-1} ( g^{\mu \nu}  q_i^{\lambda}
 q_j^{\rho} + g^{\mu \lambda}  q_i^{\nu}  q_j^{\rho} +
    g^{\nu \lambda}   q_i^{\mu}  q_j^{\rho} +
 g^{\mu \rho}   q_i^{\nu}
 q_j^{\lambda}   
+g^{\nu \rho}   q_i^{\mu}  q_j^{\lambda} +
     g^{\lambda \rho}   q_i^{\mu}  q_j^{\nu}  ) n_{ij} I_{n,ij}^{[d+]^3}  
\nn\\
&&  
+~\frac{1}{4} \left(\, g^{\mu \nu}  \,g^{\lambda \rho}  \,+ g^{\mu \lambda}  \,  g^{\nu \rho} \,+
    \, g^{\mu \rho}  \,  g^{\nu \lambda} \, \right)I_{n}^{[d+]^2} \, ,
\label{tensorred1}
\end{eqnarray}
where  $[d+]$ is an operator shifting the space-time dimension by two units and
\begin{eqnarray}
  \label{eq:Inij}
   I_{p, \, i\,j \,k\cdots} ^{[d+]^l,stu \cdots} &=&  \int ^{[d+]^l}  \prod_{r=1}^{n} \, \frac{1}{c_r^{1+\delta_{ri} + \delta_{rj}+\delta_{rk}+\cdots
                -\delta_{rs} - \delta_{rt}-\delta_{ru}-\cdots}},
\nl
   \int ^{d} &\equiv&  \int \frac{d^d k}{{i\pi}^{d/2}},
\end{eqnarray}
where $[d+]^l=4+2 l -2 \eps$ (observe that $p$ is the number of scalar
propagators of the ``$p$-point function'' and that equal lower and upper indices
 cancel, $p\leq n$).
In (\ref{vectorred}--\ref{tensorred1}), the coefficients
$n_{ij}, n_{ijk}$ and $n_{ijkl}$ were introduced. 
These stand for the product of factorials of the number
of equal indices: e.g. $n_{iiii}=4!, n_{ijii}=3!, n_{iijj}=2! 2!, n_{ijkk}=2!,n_{ijkl}=1! $
(indices $i,j,k,l$ all different from each other). Of particular relevance are the following relations
for the successive application of recurrence relations to reduce higher dimensional integrals:
\begin{eqnarray}
 n_{ij}&=&{\nu}_{ij},
\nl
n_{ijk}&=&{\nu}_{ij} {\nu}_{ijk},
\nl
n_{ijkl}&=&{\nu}_{ij} {\nu}_{ijk} {\nu}_{ijkl},
\end{eqnarray}
where
\begin{eqnarray}
{\nu}_{ij}&=&1+{\delta}_{ij} ,
\nl
{\nu}_{ijk}&=&1+{\delta}_{ik}+{\delta}_{jk},
\nl
{\nu}_{ijkl}&=&1+{\delta}_{il}+{\delta}_{jl}+{\delta}_{kl}.
\end{eqnarray}
In the next step the integrals in higher dimension have to be reduced to integrals in generic dimension.
Here particular attention has to be paid to  $I_{5}^{[d+]}$. Reducing the
tensor integrals, this term drops out in general
 \cite{Bern:1994kr,Binoth:1999sp}. 

\section {\label{rank1234}Pentagons}
We start with the reduction of the pentagons. 
This will also provide the basis for calculating the hexagons as we shall see.

\subsection{\label{rank01}Scalar and vector integrals}
For the {\bf scalar} 5-point function the recursion relation (I.31)
reads for $n=5$
\begin{eqnarray}\label{i5sc}
(d-4) {\begin{pmatrix} ~\\ ~\end{pmatrix} }_5 I_{5}^{[d+]}={\begin{pmatrix} 0\\ 0\end{pmatrix} }_5 I_{5}
-\sum_{s=1}^{5} \begin{pmatrix} 0\\ s\end{pmatrix} _5 I_{4}^{s}
\label{scalargn}
\end{eqnarray}
With  $I_{5}^{[d+]}$ finite for $d=4$, we have in this limit
\begin{eqnarray}\label{i5sc2}
E \equiv I_{5}
&=&          
\frac{1}
{
           \begin{pmatrix} 0 \\ 0\end{pmatrix}_5
}
\sum_{s=1}^{5} {\begin{pmatrix} 0 \\ s \end{pmatrix} }_5 
I_{4}^{s},
\label{scalar4p}
\end{eqnarray}
i.e. the scalar five-point function is expressed in the limit $d \to 4$
in terms of scalar four-point functions, which are obtained by
scratching in the five terms of the sum the $s^{th}$ scalar propagator,
respectively.
This was already derived in \cite{Melrose:1965kb}, see eq. (6.1) there.

Similarly, for the tensor integral
of rank $R=1$ (vector) in (\ref{scalarred})  we obtain:
\begin{eqnarray}\label{i5vc1}
 I_{5}^{\mu}
&=&
\sum_{i=1}^{4} \, q_i^{\mu} I_{5,i},
\end{eqnarray}
with
\begin{eqnarray}  \label{i5vc2}
I_{5,i} \equiv E_i
&=&
-I_{5,i}^{[d+]} 
\nl
&=&
 (d-4) 
\frac{
\begin{pmatrix} 0 \\ i \end{pmatrix}_5
}
{
\begin{pmatrix} 0 \\ 0 \end{pmatrix}_5
} 
I_{5}^{[d+]}-
\frac{1}{
\begin{pmatrix} 0 \\ 0 \end{pmatrix}_5
} 
\sum_{s=1}^{5} \begin{pmatrix} 0&i \\ 0&s \end{pmatrix}_5 
I_{4}^{s},
\label{first}
\end{eqnarray}
where again in the limit $d \to 4$ the  $I_{5}^{[d+]}$
disappears.\footnote{The $I_{5,i}$ should not be confused with quantities introduced in Equation (\ref{eq:Inij}).}
These two cases are simple and lead to a direct
reduction to scalar integrals, without the Gram determinant $()_5$
occurring anyway. In the following we want to reduce tensor integrals
of higher rank
and show, like in \cite{Denner:2002ii,Binoth:2005ff}, that also in these cases
the Gram determinant can be cancelled.

\subsection{\label{rank2}$R=2$ tensor integrals}
The tensor integral of {\bf rank 2} can be written without a $g_{\mu\nu}$-term:
\be
 I_{5}^{\mu\, \nu}= \sum_{i,j=1}^{4} \, q_i^{\mu}\, q_j^{\nu}  I_{5,ij},
\label{tensor2}
\ee
which is obtained by replacing $g_{\mu \nu}$ by
\be
g^{\mu\, \nu}=2 \sum_{i,j=1}^{4} 
\frac{
{\begin{pmatrix} i \\ j \end{pmatrix} }_5
}
{
{\begin{pmatrix} ~\\ ~\end{pmatrix} }_5
} 
\, q_i^{\mu}\, q_j^{\nu}
\label{gmunu}
\ee
(assuming $q_1 \cdots q_4$ 4-dimensional and independent)
and further by reducing the integrals in (\ref{vectorred}) to generic
dimension. This applies in the same manner also for the tensor integrals
of higher rank. Reducing the integrals of highest dimension, recursion
relation (I.30) is used.
For the 5-point function several cases have
been worked out in : (I.41), (I.42) and (I.43). For
completeness we give in the Appendix explicitly  the cases needed in
the present work.

One remark is needed concerning the integral $I_{5}^{[d+]}$. It is known
\cite{Bern:1994kr,Binoth:1999sp} that it always cancels in the end.
This provides a very useful check on our calculations,
which we have performed in every particular case under consideration. 
Anticipating this cancellation, we will,
for the ease of our discussion, drop terms proportional to $I_{5}^{[d+]}$
wherever they appear in the following derivation.
With this in mind we can write for $I_{5,ij}$ in (\ref{vectorred})
with (\ref{A522}):
\bea
I_{5,ij}&=&{\nu}_{ij} I_{5,ij}^{[d+]^2}
\nl
&=&-\frac{
{\begin{pmatrix} 0 \\ j \end{pmatrix} }_5}{{\begin{pmatrix} ~\\ ~\end{pmatrix} }_5} I_{5,i}^{[d+]} +
 \sum_{s=1,s \ne i}^{5} \frac{{
\begin{pmatrix} s \\ j \end{pmatrix} }_5}{{\begin{pmatrix} ~\\ ~\end{pmatrix} }_5} I_{4,i}^{[d+],s}
\nl &=&
\frac{{
\begin{pmatrix} 0 \\ j \end{pmatrix} }_5}{{\begin{pmatrix} ~\\ ~\end{pmatrix} }_5} I_{5,i}+ \sum_{s=1,s \ne i}^{5} \frac{{
\begin{pmatrix} s \\ j \end{pmatrix} }_5}
{{\begin{pmatrix} ~\\ ~\end{pmatrix} }_5} I_{4,i}^{[d+],s} ,
\label{tenstwo}
\eea
and by means of (\ref{A411}) we obtain:
\bea
I_{5,ij}
=&&
\frac{1}
{
{\begin{pmatrix} 0 \\ 0 \end{pmatrix} }_5 
{\begin{pmatrix} ~\\ ~\end{pmatrix} }_5
}
\sum_{s=1,s \ne i}^{5} 
\frac{1}
{
{\begin{pmatrix} s \\ s \end{pmatrix} }_5
}
\left\{
- 
{\begin{pmatrix} 0     \\ j  \end{pmatrix} }_5 
{\begin{pmatrix} 0&  s \\ 0&i\end{pmatrix} }_5 
{\begin{pmatrix} s     \\ s  \end{pmatrix} }_5 
-
{\begin{pmatrix} s     \\ j   \end{pmatrix} }_5 
{\begin{pmatrix} 0&  s \\ i&s \end{pmatrix} }_5 
{\begin{pmatrix} 0     \\ 0   \end{pmatrix} }_5 
\right. 
\nn \\
&&
+~\left.
{\begin{pmatrix} s     \\ 0   \end{pmatrix} }_5  
{\begin{pmatrix} 0&  s \\ 0&s \end{pmatrix} }_5 
{\begin{pmatrix} i      \\ j  \end{pmatrix} }_5
\right\}
I_4^s
-
\frac{
{\begin{pmatrix} i \\ j \end{pmatrix} }_5
}
{
{\begin{pmatrix} 0 \\ 0 \end{pmatrix} }_5 
{\begin{pmatrix} ~\\ ~\end{pmatrix} }_5
}
\sum_{s=1,s \ne i}^{5} 
\frac{1}{
{\begin{pmatrix} s \\ s \end{pmatrix} }_5
}
{\begin{pmatrix} s \\ 0 \end{pmatrix} }_5  
{\begin{pmatrix} 0&  s \\ 0&s \end{pmatrix} }_5 
I_4^s
\nn \\
&&
- ~\frac{1}
{
{\begin{pmatrix} 0 \\ 0 \end{pmatrix} }_5 
{\begin{pmatrix} ~\\ ~\end{pmatrix} }_5
}
\sum_{s,t=1,s \ne i,t}^{5} \frac{1}
{
{\begin{pmatrix} s \\ s \end{pmatrix} }_5
}
\left\{
- 
{\begin{pmatrix} 0     \\ j \end{pmatrix} }_5 
{\begin{pmatrix} t&  s \\0&i\end{pmatrix} }_5 
{\begin{pmatrix} s     \\ s \end{pmatrix} }_5
-
{\begin{pmatrix} s     \\ j \end{pmatrix} }_5 
{\begin{pmatrix} t&  s \\i&s\end{pmatrix} }_5 
{\begin{pmatrix} 0     \\ 0 \end{pmatrix} }_5 
\right.
\nn \\
&&
 +~\left. 
{\begin{pmatrix} s     \\ 0   \end{pmatrix} }_5  
{\begin{pmatrix} t&  s \\ 0&s \end{pmatrix} }_5 
{\begin{pmatrix} i     \\ j   \end{pmatrix} }_5
\right\}
I_3^{st}
+ 
\frac{
{\begin{pmatrix} i \\ j \end{pmatrix} }_5
}
{
{\begin{pmatrix} 0 \\ 0 \end{pmatrix} }_5 
{\begin{pmatrix} ~\\ ~\end{pmatrix} }_5
}
\sum_{s,t=1,s \ne i,t}^{5} 
\frac{1}
{
{\begin{pmatrix} s \\ s \end{pmatrix} }_5
}
{\begin{pmatrix} s     \\ 0   \end{pmatrix} }_5  
{\begin{pmatrix} t&  s \\ 0&s \end{pmatrix} }_5 
I_3^{st}
.
\label{inter2}
\eea
Using (\ref{gmunu}) again, we find
\begin{eqnarray}
I_{5}^{\mu\, \nu\,}&=& \sum_{i,j=1}^{4} \, q_i^{\mu}\, q_j^{\nu} E_{ij} +
g^{\mu \nu}  E_{00}, 
\\
E_{ij}&=&\sum_{s=1}^{5} S_{ij}^{4,s} I_4^s +\sum_{s,t=1}^{5} S_{ij}^{3,st} I_3^{st} ,
\label{final2}
\end{eqnarray}
where
\begin{eqnarray}
S_{ij}^{4,s}&=&\frac{1}{{\begin{pmatrix} 0 \\ 0 \end{pmatrix} }_5} \sum_{s=1}^{5} \frac{1}{{\begin{pmatrix} s \\ s \end{pmatrix} }_5} X_{ij}^{s0}, 
\\
S_{ij}^{3,st}&=&-\frac{1}{{\begin{pmatrix} 0 \\ 0 \end{pmatrix} }_5} \sum_{s,t=1}^{5} \frac{1}{{\begin{pmatrix} s \\ s \end{pmatrix} }_5} X_{ij}^{st}
\end{eqnarray}
and $X_{ij}^{s0}$ and $X_{ij}^{st}$ are defined in (\ref{xijst}).
Finally,
\bea
E_{00}&=&-\frac{1}{2} 
\frac{1}{
{\begin{pmatrix} 0 \\ 0 \end{pmatrix} }_5
} \sum_{s=1}^5
\frac{
{\begin{pmatrix} s \\ 0 \end{pmatrix} }_5
}
{
{\begin{pmatrix} s \\ s \end{pmatrix} }_5
}
\left[ 
{\begin{pmatrix} 0&  s \\ 0&s \end{pmatrix} }_5 
I_4^s 
- 
\sum_{t=1}^{5} 
{\begin{pmatrix} t&  s \\ 0&s \end{pmatrix} }_5 
I_3^{st} \right] \, .
\eea
In this way we have cancelled the Gram determinant for the tensor of rank 2.
For later reference, we note that, by taking into account (\ref{A401}),
we can also write
\bea
E_{00}&=&
-\frac{1}{2} 
\frac{1}
{
{\begin{pmatrix} 0 \\ 0 \end{pmatrix} }_5
} \sum_{s=1}^5
{\begin{pmatrix} s \\ 0 \end{pmatrix} }_5 
I_4^{[d+],s}.
\label{E00a}
\eea

\subsection{\label{appT3}$R=3$ tensor integrals}
The tensor integral of {\bf rank 3} can be written as:
\begin{eqnarray}
I_{5}^{\mu\, \nu\, \lambda}
= \sum_{i,j,k=1}^{4} \, q_i^{\mu}\, q_j^{\nu} \, q_k^{\lambda}  I_{5,ijk}.
\label{tensor3}
\end{eqnarray}
We will now rewrite this into another representation, thereby avoiding 
Gram determinants $\left(  \right)_5$
in the denominators of the new tensor coefficients $E_{ijk}, E_{00k}$:  
\begin{eqnarray}\label{final3c}
I_{5}^{\mu\, \nu\, \lambda}
&=& 
\sum_{i,j,k=1}^{4} \, q_i^{\mu}\, q_j^{\nu} \, q_k^{\lambda}
E_{ijk}+\sum_{k=1}^4 g^{[\mu \nu} q_k^{\lambda]} E_{00k},
\\ 
E_{ijk}&=&\sum_{s=1}^{5} S_{ijk}^{4,s} I_4^s +\sum_{s,t=1}^{5} S_{ijk}^{3,st} I_3^{st}+
\sum_{s,t,u=1}^{5} S_{ijk}^{2,stu} I_2^{stu} .
\label{final3}
\end{eqnarray}
According to (\ref{intone}) we have with (\ref{gmunu}):
\begin{eqnarray}
 I_{5,ijk}
=-~{\nu}_{ij} {\nu}_{ijk} I_{5,ijk}^{[d+]^3}
+
\frac
{
{\begin{pmatrix} j \\ k \end{pmatrix} }_5
}
{
{\begin{pmatrix} ~\\ ~\end{pmatrix} }_5
} 
I_{5,i}^{[d+]^2}
+
\frac{
{\begin{pmatrix} i \\ k \end{pmatrix} }_5
}
{
{\begin{pmatrix} ~\\ ~\end{pmatrix} }_5} 
I_{5,j}^{[d+]^2}
+
\frac{
{\begin{pmatrix} i \\ j \end{pmatrix} }_5
}
{
{\begin{pmatrix} ~\\ ~\end{pmatrix} }_5} 
I_{5,k}^{[d+]^2} 
.
\label{raw3}
\end{eqnarray}
By means of recursion (\ref{A533}), taking into account (\ref{tenstwo}) and keeping in mind to drop
$I_5^{[d+]}$, we have:
\begin{eqnarray}
 I_{5,ijk}&&=\frac{{\begin{pmatrix} 0 \\ k \end{pmatrix} }_5}{{\begin{pmatrix} ~\\ ~\end{pmatrix} }_5} {\nu}_{ij} I_{5,ij}^{[d+]^2}-
\sum_{s=1,s \ne i,j}^{5} \frac{ {\begin{pmatrix} s \\ k \end{pmatrix} }_5}{{\begin{pmatrix} ~\\ ~\end{pmatrix} }_5}  {\nu}_{ij} I_{4,ij}^{[d+]^2,s}+
\frac{ {\begin{pmatrix} i \\ j \end{pmatrix} }_5}{{\begin{pmatrix} ~\\ ~\end{pmatrix} }_5} \sum_{s=1}^{5} \frac{ {\begin{pmatrix} s \\ k \end{pmatrix} }_5}{{\begin{pmatrix} ~\\ ~\end{pmatrix} }_5}
I_4^{[d+],s} \nn\\
&&=\frac{{\begin{pmatrix} 0 \\ k \end{pmatrix} }_5}{{\begin{pmatrix} ~\\ ~\end{pmatrix} }_5} I_{5,ij}+\frac{{\begin{pmatrix} i \\ j \end{pmatrix} }_5}{{\begin{pmatrix} ~\\ ~\end{pmatrix} }_5}
\sum_{s=1}^{5} \frac{ {\begin{pmatrix} s \\ k \end{pmatrix} }_5}{{\begin{pmatrix} ~\\ ~\end{pmatrix} }_5} I_4^{[d+],s}-
\sum_{s=1,s \ne i,j}^{5} \frac{ {\begin{pmatrix} s \\ k \end{pmatrix} }_5}{{\begin{pmatrix} ~\\ ~\end{pmatrix} }_5}  {\nu}_{ij} I_{4,ij}^{[d+]^2,s} .
\label{refine3}
\end{eqnarray}
Collecting the terms proportional to $\left(  \right)_5^{-2}$ we have with
$I_{5,ij}= \cdots +2\frac{{\begin{pmatrix} i \\ j \end{pmatrix} }_5}{{\begin{pmatrix} ~\\ ~\end{pmatrix} }_5} E_{00}$ and (\ref{E00a}):
\begin{eqnarray}
\frac{
{\begin{pmatrix} i \\ j \end{pmatrix} }_5
}
{
{\begin{pmatrix} 0 \\ 0 \end{pmatrix} }_5
} 
\frac{1}
{
{\begin{pmatrix} ~\\ ~\end{pmatrix} }_5^2
} 
\sum_{s=1}^{5}
\left[
{\begin{pmatrix} 0 \\ 0 \end{pmatrix} }_5 
{\begin{pmatrix} s \\ k \end{pmatrix} }_5
-
{\begin{pmatrix} 0 \\ k \end{pmatrix} }_5 
{\begin{pmatrix} s \\ 0 \end{pmatrix} }_5 
\right]  
I_4^{[d+],s}
=
\frac{
{\begin{pmatrix} i \\ j \end{pmatrix} }_5
}
{
{\begin{pmatrix} 0 \\ 0 \end{pmatrix} }_5}
 \frac{1}
{
{\begin{pmatrix} ~\\ ~\end{pmatrix} }_5} 
\sum_{s=1}^{5}
{\begin{pmatrix} 0&  s \\ 0&k \end{pmatrix} }_5 
I_4^{[d+],s},
\nl
\end{eqnarray}
i.e. we have already cancelled one Gram determinant.
We multiply (\ref{refine3}) by ${\begin{pmatrix} 0 \\ 0 \end{pmatrix} }_5$ such that we can make use of
\begin{eqnarray}
{\begin{pmatrix} 0 \\ 0 \end{pmatrix} }_5  
{\begin{pmatrix} s \\ k \end{pmatrix} }_5 
&=& 
{\begin{pmatrix} 0&  s \\ 0&k \end{pmatrix} }_5 
{\begin{pmatrix} ~\\ ~\end{pmatrix} }_5 
+ 
{\begin{pmatrix} s \\ 0 \end{pmatrix} }_5 {\begin{pmatrix} 0 \\ k \end{pmatrix} }_5,
\label{zzsk}
\end{eqnarray}
which will give us another factor $\left(  \right)_5$.
Adding all contributions, we obtain
\begin{eqnarray}
3 
{\begin{pmatrix} 0 \\ 0 \end{pmatrix} }_5 I_{5,ijk}
&=&
{\sum}^{'}
\frac{1}{{\begin{pmatrix} ~\\ ~\end{pmatrix} }_5  {\begin{pmatrix} s \\ s \end{pmatrix} }_5^2} 
\Biggl\{
{\begin{pmatrix} 0 \\ k \end{pmatrix} }_5 
{\begin{pmatrix} s \\ s \end{pmatrix} }_5 
\left[ X_{ij}^s I_4^s - X_{ij}^{st} I_3^{st} \right]
\nl 
&&+~ 
{\begin{pmatrix} i \\ j \end{pmatrix} }_5 
{\begin{pmatrix} s \\ s \end{pmatrix} }_5 
{\begin{pmatrix} 0&  s \\ 0&k \end{pmatrix} }_5 
\left[
{\begin{pmatrix} 0&  s \\ 0&s \end{pmatrix} }_5 
I_4^s -  {\begin{pmatrix} t&  s \\ 0&s \end{pmatrix} }_5
I_3^{st} \right] 
\nl
&&-~ {\begin{pmatrix} 0 \\ 0 \end{pmatrix} }_5 {\begin{pmatrix} s \\ k \end{pmatrix} }_5 
\Biggl( \left[
{\begin{pmatrix} 0&  s \\ i&s \end{pmatrix} }_5 {\begin{pmatrix} 0&  s \\ j&s\end{pmatrix} }_5+{\begin{pmatrix} i&  s \\ j&s\end{pmatrix} }_5{\begin{pmatrix} 0&  s \\ 0&s \end{pmatrix} }_5\right]  
I_{4}^{s}
\nl
&&-~
                \left[{\begin{pmatrix} 0&  s \\ j&s\end{pmatrix} }_5 {\begin{pmatrix} t&  s \\ i&s \end{pmatrix} }_5+{\begin{pmatrix} i&  s \\ j&s\end{pmatrix} }_5{\begin{pmatrix} t&  s \\ 0&s \end{pmatrix} }_5 \right] I_{3}^{st}
         \Biggr)
\Biggr\}  
\nn\\
&&
{}+{\sum}^{'}
\frac{1}{ {\begin{pmatrix} ~\\ ~\end{pmatrix} }_5 } 
\frac{
{\begin{pmatrix} 0 \\ 0 \end{pmatrix} }_5 
{\begin{pmatrix} s \\ k \end{pmatrix} }_5 
} 
{
{\begin{pmatrix} s \\ s \end{pmatrix} }_5 
{\begin{pmatrix} s&t \\ s&t \end{pmatrix} }_5
}
{\begin{pmatrix} t&  s \\ j&s\end{pmatrix} }_5
\left[
{\begin{pmatrix} 0&s&t  \\ i&s&t\end{pmatrix} }_5 
I_{3}^{st}
-
{\begin{pmatrix} u&s&t  \\ i&s&t\end{pmatrix} }_5 
I_{2}^{stu} 
\right]  
\nl
&&+~ 
(i \leftrightarrow k) + (j \leftrightarrow k)
\nl
&\equiv&
A ~~ + 
{\sum}^{'}
\Biggl\{
\frac{1}{{\begin{pmatrix} ~\\ ~\end{pmatrix} }_5  {\begin{pmatrix} s \\ s \end{pmatrix} }_5^2} 
{\begin{pmatrix} i \\ j \end{pmatrix} }_5 
{\begin{pmatrix} s \\ s \end{pmatrix} }_5 
{\begin{pmatrix} 0&  s \\ 0&k \end{pmatrix} }_5 
\left[
{\begin{pmatrix} 0&  s \\ 0&s \end{pmatrix} }_5 
I_4^s -  {\begin{pmatrix} t&  s \\ 0&s \end{pmatrix} }_5
I_3^{st} \right] 
\nl
&& 
+~
\frac{1}{ {\begin{pmatrix} ~\\ ~\end{pmatrix} }_5 } 
\frac{
{\begin{pmatrix} 0 \\ 0 \end{pmatrix} }_5 
{\begin{pmatrix} s \\ k \end{pmatrix} }_5 
} 
{
{\begin{pmatrix} s \\ s \end{pmatrix} }_5 
{\begin{pmatrix} s&t \\ s&t \end{pmatrix} }_5
}
{\begin{pmatrix} t&  s \\ j&s\end{pmatrix} }_5
\left[
{\begin{pmatrix} 0&s&t  \\ i&s&t\end{pmatrix} }_5 
I_{3}^{st}
-
{\begin{pmatrix} u&s&t  \\ i&s&t\end{pmatrix} }_5 
I_{2}^{stu} 
\right] \Biggr\}
\nl
&&+~ 
(i \leftrightarrow k) + (j \leftrightarrow k)
\label{addings}
\end{eqnarray}
The symbol ${\sum}^{'}$ in these equations denotes
a sum
$\sum_{s,t,u=1}^{5}$ in terms proportional to $I_2^{stu}$,
$\sum_{s,t=1}^{5}$ in terms proportional to $I_3^{st}$,
and
$\sum_{s=1}^{5}$ in terms proportional to $I_4^{s}$.
Concerning the symmetrization in (\ref{addings}), we point out that
the original expression (\ref{raw3}) is obviously symmetric under
$(i \leftrightarrow j)$, while this is not explicitly seen in (\ref{addings})
anymore. Later on, however, this symmetry will become apparent again.

All terms with factors of the type
${\begin{pmatrix} i \\ j \end{pmatrix} }_5$ can be considered,
due to (\ref{gmunu}), as belonging to some
$g_{\mu \nu}$ term.  For other terms we have to use (\ref{zzsk}),
which yields terms with $\left(  \right)_5$ to be cancelled. These are
explicitly given in the coefficients of $I_4^s, I_3^{st}$ and $I_2^{stu}$,
i.e. (\ref{S4}, \ref{S3}, \ref{S2}).  Apart from the terms in the last
line of (\ref{addings}) and the
${\begin{pmatrix} i \\ j \end{pmatrix} }_5$ term,
the remaining contributions to the coefficients of  $I_4^s$
and $I_3^{st}$, inserting $X_{ij}^s$ and $X_{ij}^{st}$, can be written as
\begin{eqnarray}
A &=&
- \sum_{s=1}^{5}
\frac{1}
{
{\begin{pmatrix} ~\\ ~\end{pmatrix} }_5  {\begin{pmatrix} s \\ s \end{pmatrix} }_5^2
} 
{\begin{pmatrix} 0 \\ k \end{pmatrix} }_5 
\Biggl\{ 
{\begin{pmatrix} s \\ s \end{pmatrix} }_5 
\left[ 
{\begin{pmatrix} 0&  s \\0&i \end{pmatrix} }_5 
{\begin{pmatrix} 0&  s \\ j&s\end{pmatrix} }_5
- 
{\begin{pmatrix}0&j \\  s&i \end{pmatrix} }_5
{\begin{pmatrix} 0&  s \\ 0&s \end{pmatrix} }_5 
\right]
\nl
&&+~
{\begin{pmatrix} s \\ 0 \end{pmatrix} }_5 
\left[ 
{\begin{pmatrix} 0&  s \\ i&s \end{pmatrix} }_5 
{\begin{pmatrix} 0&  s \\ j&s \end{pmatrix} }_5
+ 
{\begin{pmatrix} i&  s \\ j&s \end{pmatrix} }_5 
{\begin{pmatrix} 0&  s \\ 0&s \end{pmatrix} }_5 
\right]
\Biggr\} 
I_4^s
\nn\\
&&+~
\sum_{s,t=1}^{5}
\frac{1}
{
{\begin{pmatrix} ~\\ ~\end{pmatrix} }_5  
{\begin{pmatrix} s \\ s \end{pmatrix} }_5^2
} 
{\begin{pmatrix} 0 \\ k \end{pmatrix} }_5 
\Biggl\{ 
{\begin{pmatrix} s     \\ s   \end{pmatrix} }_5 
\left[ 
{\begin{pmatrix} 0&  s \\ 0&j \end{pmatrix} }_5 
{\begin{pmatrix} t&  s \\ i&s \end{pmatrix} }_5
- 
{\begin{pmatrix} 0&i   \\ s&j \end{pmatrix} }_5 
{\begin{pmatrix} t&  s \\ 0&s \end{pmatrix} }_5 
\right]
\nl
&&+~
{\begin{pmatrix} s     \\ 0   \end{pmatrix} }_5 
\left[ 
{\begin{pmatrix} 0&  s \\ j&s \end{pmatrix} }_5 
{\begin{pmatrix} t&  s \\ i&s \end{pmatrix} }_5
+ 
{\begin{pmatrix} j&s   \\ i&s \end{pmatrix} }_5 
{\begin{pmatrix} t&  s \\ 0&s \end{pmatrix} }_5 
\right]
\Biggr\} 
I_3^{st}. 
\label{I43}
\end{eqnarray}
Here the following  ``master formula'' ( Equation (A.13) of \cite{Melrose:1965kb} ) is of great help:
\begin{eqnarray}
{\begin{pmatrix} s \\ i \end{pmatrix} }_5 
{\begin{pmatrix} s &{\tau}\\ 0&s \end{pmatrix} }_5=
{\begin{pmatrix} s \\ 0 \end{pmatrix} }_5{\begin{pmatrix} s& {\tau}\\ i&s \end{pmatrix} }_5+{\begin{pmatrix} s \\ s \end{pmatrix} }_5
 {\begin{pmatrix} s &{\tau}\\ 0&i \end{pmatrix}}_5,~~~ \tau = 0,1, \dots 5,
\label{Zauber}
\end{eqnarray}
which yields explicitly:
\begin{eqnarray}
{\begin{pmatrix} s \\ s \end{pmatrix} }_5 
{\begin{pmatrix} 0&  s \\ 0&i\end{pmatrix} }_5
+
{\begin{pmatrix} s \\ 0 \end{pmatrix} }_5
{\begin{pmatrix} 0&  s \\ i&s \end{pmatrix} }_5
&=&~~~
 {\begin{pmatrix} s \\ i \end{pmatrix} }_5
{\begin{pmatrix} 0&  s \\ 0&s \end{pmatrix} }_5 ,
\end{eqnarray}
and: 
\begin{eqnarray}
{\begin{pmatrix} s \\ s \end{pmatrix} }_5 
{\begin{pmatrix}0&j \\ s&i \end{pmatrix} }_5
-
{\begin{pmatrix} s \\ 0 \end{pmatrix} }_5
{\begin{pmatrix} i&  s \\ j&s\end{pmatrix} }_5
&=&
-
{\begin{pmatrix} s \\ j \end{pmatrix} }_5 
{\begin{pmatrix} 0&  s \\ i&s \end{pmatrix} }_5,
\label{Zauberei}
\end{eqnarray}
so that (\ref{I43}) reads:
\begin{eqnarray}
A&=&
{} - \sum_{s=1}^{5}
\frac{1}{{\begin{pmatrix} ~\\ ~\end{pmatrix} }_5  
{\begin{pmatrix} s \\ s \end{pmatrix} }_5^2} 
{\begin{pmatrix} 0 \\ k \end{pmatrix} }_5 
\cdot 
\left\{
\left[
{\begin{pmatrix} s \\ i \end{pmatrix} }_5
{\begin{pmatrix} 0&  s \\ j&s\end{pmatrix} }_5 
+
{\begin{pmatrix} s \\ j \end{pmatrix} }_5 
{\begin{pmatrix} 0&  s \\ i&s \end{pmatrix} }_5 
\right]
{\begin{pmatrix} 0&  s \\ 0&s \end{pmatrix} }_5 
I_4^s
\right. 
\nn \\
&&
{} - \sum_{t=1}^{5}
\left.
\left[
{\begin{pmatrix} s \\ j \end{pmatrix} }_5
 {\begin{pmatrix} 0&  s \\ 0&s \end{pmatrix} }_5
 {\begin{pmatrix} t&  s \\ i&s \end{pmatrix} }_5
+
{\begin{pmatrix} s \\ i \end{pmatrix} }_5 
{\begin{pmatrix} 0&  s \\ j&s\end{pmatrix} }_5 
{\begin{pmatrix} t&  s \\ 0&s \end{pmatrix} }_5
\right] I_3^{st}
\label{I43p}
\right\} .
\end{eqnarray}
Next we will use :
\begin{eqnarray}
{\begin{pmatrix} 0 \\ k \end{pmatrix} }_5 {\begin{pmatrix} s \\ i \end{pmatrix} }_5 = -{\begin{pmatrix} 0&i \\ s&k \end{pmatrix} }_5 {\begin{pmatrix} ~\\ ~\end{pmatrix} }_5 + {\begin{pmatrix} i \\ k \end{pmatrix} }_5 {\begin{pmatrix} s \\ 0 \end{pmatrix} }_5 .
\label{product}
\end{eqnarray}
As trivial as this relation may look, it plays the crucial role of splitting off
${\begin{pmatrix} i \\ k \end{pmatrix} }_5$ in order to produce $g^{\mu \nu}$ terms. 
It might also have been written as:
\begin{eqnarray}
{\begin{pmatrix} 0 \\ k \end{pmatrix} }_5 {\begin{pmatrix} s \\ i \end{pmatrix} }_5 
&=&
 {\begin{pmatrix} 0&  s \\  k&i\end{pmatrix}}_5 {\begin{pmatrix} ~\\ ~\end{pmatrix} }_5 +
{\begin{pmatrix} 0 \\ i \end{pmatrix} }_5
{\begin{pmatrix} s \\ k \end{pmatrix} }_5 ,
\end{eqnarray}
but then it would not fulfill its purpose.

The first term at the rhs. of (\ref{product}) cancels a $\left(  \right)_5$,
while the second term enters the $g_{\mu \nu}$-terms,
all of which are collected in (\ref{E00j}).
The complete coefficient of $I_4^s$  in (\ref{final3c}) is thus given by:
\begin{eqnarray}
S_{ijk}^{4,s}&=&
\frac{1}
{3 
{\begin{pmatrix} 0 \\ 0 \end{pmatrix} }_5  
{\begin{pmatrix} s \\ s \end{pmatrix} }_5^2}
\Biggl\{
-
{\begin{pmatrix} 0&  s \\ 0&k \end{pmatrix} }_5 
\left[
{\begin{pmatrix} 0&  s \\ i&s \end{pmatrix} }_5 
{\begin{pmatrix} 0&  s \\ j&s\end{pmatrix} }_5 +
{\begin{pmatrix} i&  s \\ j&s\end{pmatrix} }_5 
{\begin{pmatrix} 0&  s \\ 0&s \end{pmatrix} }_5 
\right]
\nl
&&+~
 \left[
{\begin{pmatrix} 0&i \\ s&k \end{pmatrix} }_5 
{\begin{pmatrix} 0&  s \\ j&s\end{pmatrix} }_5
+
{\begin{pmatrix}0&j \\  s&k \end{pmatrix} }_5 
{\begin{pmatrix} 0&  s \\ i&s \end{pmatrix} }_5 
\right]
{\begin{pmatrix} 0&  s \\ 0&s \end{pmatrix} }_5 
+ (i \leftrightarrow k) + (j \leftrightarrow k) 
\Biggr\} .
\label{S4}
\end{eqnarray}
Finally we have to investigate the last line of (\ref{addings}), being left with the factor
${\begin{pmatrix} 0 \\ k \end{pmatrix} }_5 {\begin{pmatrix} s \\ 0 \end{pmatrix} }_5$ as before in (\ref{zzsk}). 
The master formula (\ref{Zauber}) then yields:
\begin{eqnarray}
{\begin{pmatrix} s \\ 0 \end{pmatrix} }_5  
{\begin{pmatrix} t&  s \\ j&s\end{pmatrix} }_5
=
{\begin{pmatrix} s \\ j \end{pmatrix} }_5 
{\begin{pmatrix} t&  s \\ 0&s \end{pmatrix} }_5 - 
{\begin{pmatrix} s \\ s \end{pmatrix} }_5 
{\begin{pmatrix} t&  s \\ 0&j \end{pmatrix} }_5.
\label{miracle}
\end{eqnarray}
The ${\begin{pmatrix} s \\ s \end{pmatrix} }_5$ in the second term of (\ref{miracle}) cancels and the remaining factor is
antisymmetric in $s$ and $t$, i.e. this term drops out after summation over $s,t$. Using again
(\ref{product}) and dropping for the time being the contribution to $g_{\mu \nu}$ terms, we
finally write the coefficients of $I_3^{st}$ and $I_2^{stu}$ in the
following way, taking
care of the original $(i \leftrightarrow j)$ symmetry in (\ref{addings}):
\begin{eqnarray}
S_{ijk}^{3,st}
&=&
\frac{1}
{
3 
{\begin{pmatrix} 0 \\ 0 \end{pmatrix} }_5  
{\begin{pmatrix} s \\ s \end{pmatrix} }_5^2 
}
\Biggl\{ 
{\begin{pmatrix} 0&  s \\ 0&k \end{pmatrix} }_5
\Biggl[ 
{\begin{pmatrix} t&  s \\ i&s \end{pmatrix} }_5 
{\begin{pmatrix} 0&  s \\ j&s\end{pmatrix} }_5 
+ 
{\begin{pmatrix} i&  s \\ j&s\end{pmatrix} }_5 
{\begin{pmatrix} t&  s \\ 0&s \end{pmatrix} }_5
\nl 
&+&
\frac{
{\begin{pmatrix} s \\ s \end{pmatrix} }_5 
{\begin{pmatrix}0&s& t&  \\ i&s&t\end{pmatrix} }_5
}
{
{\begin{pmatrix} s&t \\ s&t \end{pmatrix} }_5
} 
{\begin{pmatrix} t&  s \\ j&s\end{pmatrix} }_5
\Biggr]
-
\Biggl[
{\begin{pmatrix} 0&i \\ s&k \end{pmatrix} }_5
{\begin{pmatrix} 0&  s \\ j&s\end{pmatrix} }_5
+
{\begin{pmatrix}0&j \\  s&k \end{pmatrix} }_5
{\begin{pmatrix} 0&  s \\ i&s \end{pmatrix} }_5
\Biggr]
{\begin{pmatrix} t&  s \\ 0&s \end{pmatrix} }_5
\nl
   &-&
\Biggl[
{\begin{pmatrix} 0&i \\ s&k \end{pmatrix} }_5
{\begin{pmatrix} t&  s \\ j&s\end{pmatrix} }_5
+
{\begin{pmatrix}0&j \\  s&k \end{pmatrix} }_5
{\begin{pmatrix} t&  s \\ i&s \end{pmatrix} }_5
\Biggr]
\frac{
{\begin{pmatrix} s \\ s \end{pmatrix} }_5
 {\begin{pmatrix}0&s& t&  \\ 0&s&t \end{pmatrix} }_5
}
{
 2 
{\begin{pmatrix} s&t \\ s&t \end{pmatrix} }_5
}
\nl
&&+ (i \leftrightarrow k) + (j \leftrightarrow k) \Biggr\} ,
\label{S3}
\end{eqnarray}
and
\begin{eqnarray}
&&S_{ijk}^{2,stu}
=
-
\frac{1}{3 
{\begin{pmatrix} 0 \\ 0 \end{pmatrix} }_5  
{\begin{pmatrix} s \\ s \end{pmatrix} }_5 
{\begin{pmatrix} s&t \\ s&t \end{pmatrix} }_5}
\left\{ 
{\begin{pmatrix} 0&s \\ 0&k \end{pmatrix} }_5  
{\begin{pmatrix}t& s \\ j&s\end{pmatrix} }_5 
{\begin{pmatrix} u&s&t\\ i&s&t\end{pmatrix} }_5
-
\right. 
\nn \\
&&\left.\frac{1}{2}
\left[
{\begin{pmatrix}0&j \\  s&k \end{pmatrix} }_5 
{\begin{pmatrix} u&s&t\\ i&s&t\end{pmatrix} }_5 
+ 
{\begin{pmatrix} 0&i \\ s&k \end{pmatrix} }_5 
{\begin{pmatrix}u&s&t \\ j&s&t \end{pmatrix} }_5
\right]  
{\begin{pmatrix} t&s \\ 0&s \end{pmatrix} }_5
+ (i \leftrightarrow k) + (j \leftrightarrow k) \right\} .
\label{S2}
\end{eqnarray}
At the end we can determine the $g_{\mu \nu}$  terms from the above by collecting all terms
containing factors of the type ${\begin{pmatrix} i \\ j \end{pmatrix} }_5$:
\begin{eqnarray}
\sum_{j=1}^4 g^{[\mu \nu} q_j^{\lambda]} E_{00j}
=
\frac{2}{{\begin{pmatrix} ~\\ ~\end{pmatrix} }_5} 
\sum_{ijk=1}^4 \left[
{\begin{pmatrix} j \\ k \end{pmatrix} }_5 
E_{00i}
+
{\begin{pmatrix} i \\ k \end{pmatrix} }_5 E_{00j}
+
{\begin{pmatrix} i \\ j \end{pmatrix} }_5 E_{00k} 
\right] q_i^{\mu} q_j^{\nu} q_k^{\lambda},
\label{gmunuqla}
\end{eqnarray}
where the square bracket means symmetrization of the included indices,
\begin{eqnarray}
g^{[\mu \nu} q_k^{\lambda]}
&=& g^{\mu \nu}  \, q_k^{\lambda} \,+ g^{\mu \lambda}  \, q_k^{\nu} \,+
    \, g^{\nu \lambda}  \, q_k^{\mu},
\end{eqnarray}
and use has been made of (\ref{gmunu}). 
Collecting all terms of type ${\begin{pmatrix} i \\ j \end{pmatrix} }_5$ 
in (\ref{addings}) we have:
\begin{eqnarray}
3 
{\begin{pmatrix} 0 \\ 0 \end{pmatrix} }_5 
E_{00j}
&=&
-\frac{1}{2 } \sum_{s=1}^5  \frac{1}{
{\begin{pmatrix} s \\ s \end{pmatrix} }_5^2}~
\left[
~2 
{\begin{pmatrix} s \\ 0 \end{pmatrix} }_5 
{\begin{pmatrix}0& s \\ j&s\end{pmatrix} }_5 
- 
{\begin{pmatrix} s \\ s \end{pmatrix} }_5 
{\begin{pmatrix}0& s \\ 0&j \end{pmatrix} }_5 
\right]
{\begin{pmatrix}0& s \\ 0&s \end{pmatrix} }_5 ~~ I_{4}^{s} 
\nn\\
&&+\frac{1}{2  } \sum_{s,t=1}^5~~
\Biggl\{ ~ \frac{1}{
{\begin{pmatrix} s \\ s \end{pmatrix} }_5^2
}~
\Biggl[ 
{\begin{pmatrix} s \\ 0 \end{pmatrix} }_5 
{\begin{pmatrix}   0&s \\ j&s\end{pmatrix} }_5 
- 
{\begin{pmatrix} s \\ s \end{pmatrix} }_5 
{\begin{pmatrix} 0&s \\ 0&j \end{pmatrix} }_5 
\Biggr]
{\begin{pmatrix} t&s \\ 0&s \end{pmatrix} }_5 
\nl
&&
+
\frac{1}{
{\begin{pmatrix} s \\ s \end{pmatrix} }_5^2
} 
{\begin{pmatrix} s \\ 0 \end{pmatrix} }_5 
{\begin{pmatrix} t&s \\ j&s\end{pmatrix} }_5 
{\begin{pmatrix} 0&s \\ 0&s \end{pmatrix} }_5
\nl
&&+~
\frac{1}{ 
{\begin{pmatrix} s \\ s \end{pmatrix} }_5 
{\begin{pmatrix} s&t \\ s&t \end{pmatrix} }_5} 
~
{\begin{pmatrix} s \\ 0 \end{pmatrix} }_5 
{\begin{pmatrix} t&s \\ 0&s \end{pmatrix} }_5 
{\begin{pmatrix} 0&s&t\\ j&s&t \end{pmatrix} }_5
\Biggr\} I_{3  }^{st} 
\nn\\ 
&&-\frac{1}{2} 
\sum_{s,t,u=1}^5  \frac{1}{ 
{\begin{pmatrix} s \\ s \end{pmatrix} }_5 
{\begin{pmatrix} s&t \\ s&t \end{pmatrix} }_5
}~~~
{\begin{pmatrix} s \\ 0 \end{pmatrix} }_5 
{\begin{pmatrix}   t&s \\ 0&s \end{pmatrix} }_5 
{\begin{pmatrix} u&s&t \\ j&s&t \end{pmatrix} }_5 
~~~ I_{2  }^{stu} .
\label{E00j}
\end{eqnarray}
The following relation can be proven by
multiplication with ${\begin{pmatrix} s \\ s \end{pmatrix} }_5$,
transforming it into the relation for an extensional of Equation (A.8)
of \cite{Melrose:1965kb}:
\begin{eqnarray}
{\begin{pmatrix} s \\ 0 \end{pmatrix} }_5 
{\begin{pmatrix} \mu &s&t\\ j&s&t \end{pmatrix} }_5
=
{\begin{pmatrix} s \\ j \end{pmatrix} }_5 
{\begin{pmatrix} \mu& s&t\\ 0&s&t \end{pmatrix} }_5
-
{ \begin{pmatrix} \mu&s \\ 0&j \end{pmatrix} }_5 
{\begin{pmatrix} s&t \\ s&t \end{pmatrix} }_5
+ 
{\begin{pmatrix}  t& s \\ 0&j \end{pmatrix} }_5 
{\begin{pmatrix} t&s \\ \mu &s\end{pmatrix} }_5,~~\mu=0,1, \cdots ,4 .
\nl
\label{agree}
\end{eqnarray}
It turns out to be useful for the simplification of the coefficients of
$I_3^{st}$ and $I_2^{stu}$ in (\ref{E00j}).
For the coefficient of $I_3^{st}$, we apply relation (\ref{agree})
with $\mu =0$. The last term on the r.h.s. of (\ref{agree}) is
combined with the term on the third line of (\ref{E00j})
using (\ref{miracle}):
\begin{eqnarray}
&&
\frac{1}{{\begin{pmatrix} s \\ s \end{pmatrix} }_5^2} 
\left\{
\frac{
{\begin{pmatrix} s \\ s \end{pmatrix} }_5
}
{
{\begin{pmatrix} s&t \\ s&t \end{pmatrix} }_5
} 
{\begin{pmatrix} t&s \\ 0&j \end{pmatrix} }_5
{\begin{pmatrix} t&s \\ 0&s \end{pmatrix} }_5^2
+
{\begin{pmatrix} s \\ 0 \end{pmatrix} }_5 
{\begin{pmatrix} t&s \\ j&s\end{pmatrix} }_5 
{\begin{pmatrix}  0&s \\ 0&s \end{pmatrix} }_5
\right\}
\nl
=&&
\frac{1}{
{\begin{pmatrix} s \\ s \end{pmatrix} }_5^2
}
{\begin{pmatrix} s \\ j \end{pmatrix} }_5
{\begin{pmatrix}t& s \\ 0&s \end{pmatrix} }_5
{\begin{pmatrix} 0&s \\ 0&s \end{pmatrix} }_5
-
\frac{
{\begin{pmatrix} t&s \\ 0&j \end{pmatrix} }_5
{\begin{pmatrix} 0&s&t\\ 0&s&t \end{pmatrix} }_5
}
{
{\begin{pmatrix} s&t \\ s&t \end{pmatrix} }_5 .
}, 
\end{eqnarray}
After summation over $s$ and $t$, the last term on the r.h.s. will vanish. 
Furthermore we apply (\ref{Zauber}) taking $\tau =0$.

For the coefficient of $I_2^{stu}$ in (\ref{E00j}) we apply relation
(\ref{agree}) with $\mu =u$. Since $I_2^{stu}$ is symmetric in $s,t$
and $u$, we consider the sum over all permutations of any fixed set of
values of $s,t$ and $u$.
We find that
\begin{eqnarray}
\sum_{permutations}
\frac{1}{ 
{\begin{pmatrix} s \\ s \end{pmatrix} }_5 
{\begin{pmatrix} s&t \\ s&t \end{pmatrix} }_5 
} 
{\begin{pmatrix} t&s \\ 0&s \end{pmatrix} }_5
\left[ 
{\begin{pmatrix}u& s \\ 0&j \end{pmatrix} }_5 
{\begin{pmatrix} s&t \\ s&t \end{pmatrix} }_5 
- 
{\begin{pmatrix}  t& s \\ 0&j \end{pmatrix} }_5 
{\begin{pmatrix} t&s \\ u&s\end{pmatrix} }_5 
\right]=0 ,
\end{eqnarray}
so that the two last terms on the r.h.s. of (\ref{agree}) can be dropped
in this case.
Thus we have:
\begin{eqnarray}
3 
{\begin{pmatrix} 0 \\ 0 \end{pmatrix} }_5 
E_{00j}
=&-&
\frac{1}{2 } \sum_{s=1}^5~~ 
\frac{1}{
{\begin{pmatrix} s \\ s \end{pmatrix} }_5^2
}~
\left[
~3 
{\begin{pmatrix} s \\ 0 \end{pmatrix} }_5 
{\begin{pmatrix} 0&s \\ j&s\end{pmatrix} }_5 
- 
{\begin{pmatrix} s \\ j \end{pmatrix} }_5 
{\begin{pmatrix} 0&s \\ 0&s \end{pmatrix} }_5 
\right]
{\begin{pmatrix} 0&  s \\ 0&s \end{pmatrix} }_5 
I_{4}^{s} 
\nn\\
&+&
\frac{1}{2  } \sum_{s,t=1}^5~~  
\frac{1}{
{\begin{pmatrix} s \\ s \end{pmatrix} }_5^2
}~
\left[
~3 
{\begin{pmatrix} s \\ 0 \end{pmatrix} }_5 
{\begin{pmatrix} 0&  s \\ j&s\end{pmatrix} }_5 
- 
{\begin{pmatrix} s \\ j \end{pmatrix} }_5
\frac{
{\begin{pmatrix} t&  s \\ 0&s \end{pmatrix} }_5^2 
}
{
{\begin{pmatrix} s&t \\ s&t \end{pmatrix} }_5} 
~\right]
{\begin{pmatrix} t&  s \\ 0&s \end{pmatrix} }_5
I_{3  }^{st} 
\nn\\
&-&
\frac{1}{2} \sum_{s,t,u=1}^5  
\frac{1}{ 
{\begin{pmatrix} s \\ s \end{pmatrix} }_5^2 
}~~
{\begin{pmatrix} s \\ j \end{pmatrix} }_5  
\frac{
{\begin{pmatrix} s \\ s \end{pmatrix} }_5 
{\begin{pmatrix} u&s&t  \\ 0&s&t \end{pmatrix} }_5
} 
{
{\begin{pmatrix} s&t \\ s&t \end{pmatrix} }_5}~~
{\begin{pmatrix} t&  s \\ 0&s \end{pmatrix} }_5 
I_{2  }^{stu} .
\label{E00f}
\end{eqnarray}
Collecting all contributions, our final result for the tensor of rank
3 can be written as:
\begin{eqnarray}
\label{final3b}
I_{5}^{\mu\, \nu\, \lambda}
&=& 
\sum_{i,j,k=1}^{4} \, q_i^{\mu}\, q_j^{\nu} \, q_k^{\lambda}
E_{ijk}+\sum_{k=1}^4 g^{[\mu \nu} q_k^{\lambda]} E_{00k},
\\
E_{ijk}&=&\sum_{s=1}^{5} S_{ijk}^{4,s} I_4^s +\sum_{s,t=1}^{5} S_{ijk}^{3,st} I_3^{st}+
\sum_{s,t,u=1}^{5} S_{ijk}^{2,stu} I_2^{stu} ,
\label{final3a}
\end{eqnarray}
and the coefficients $S_{ijk}^{4,s}, S_{ijk}^{3,st}, S_{ijk}^{2,stu}$
are given in (\ref{S4}), (\ref{S3}) and (\ref{S2}) and $E_{00k}$ in
(\ref{E00f}).

\section{\label{tens6} Hexagons}
The 6-point function has the nice property that the tensors of rank $R$ can be reduced to
a sum of six 5-point tensors of rank $R-1$. This property has also been derived in
\cite{Denner:2002ii}; an earlier demonstration of this property, however, has been
given already in \cite{Fleischer:1999hq}. The simplification in this
case is due to the fact that $()_6 \equiv 0$, which has extensively been discussed
in \cite{Fleischer:1999hq}.
Beyond that, in our approach, the above results
for the 5-point tensors can be directly used, thus reducing  the
6-point tensors of up to rank $R=4$ to scalar 4- and 3- and 2-point integrals. 
Particularly
simple results are thus obtained for the 6-point tensors using the results
of Appendix  \ref{signedminors} and Sections \ref{rank01} and \ref{rank2}. 
What was missing in  \cite{Fleischer:1999hq} is exactly this
simplification, which comes with the cancellation of the Gram determinant
${\left( \right)}_5$;
see Appendix A of that paper.
\subsection{ Scalar and vector integrals}
According to (I.33) we write (see \cite{Melrose:1965kb} and also (I.55)):
\begin{eqnarray}
I_6 &=&
\sum_{r=1}^6 \frac{
{\begin{pmatrix} 0 \\  r\end{pmatrix}}_6}{
{\begin{pmatrix} 0 \\ 0 \end{pmatrix} }_6} E^{~r}
\nl
&=&
          \sum_{r=1}^6 \frac{
{\begin{pmatrix} r\\ k \end{pmatrix} }_6}{
{\begin{pmatrix} 0 \\ k \end{pmatrix} }_6} E^{~r},~~ k=1, \dots ,6 ,
\end{eqnarray}
and  (\ref{scalar4p}) now reads:
\begin{eqnarray}
E^{~r} \equiv I_5^r=\frac{1}{
{\begin{pmatrix} 0&r\\0&r \end{pmatrix}}_6}\sum_{s=1, s \ne r}^{6} 
{\begin{pmatrix} 0&r\\s&r \end{pmatrix} }_6 I_{4}^{rs}.
\end{eqnarray}
Here we see already ~the general scheme ~of reducing ~6-point functions
~to 5-point functions: 
In general, in any signed minor $\left( \cdots \right)_5$ a further column
${\begin{pmatrix} r \\ r \end{pmatrix}}$
is  scratched, resulting in a $\left( \cdots \right)_6$
and in the scalar functions a further propagator is scratched.

As in (\ref{i5vc1}) and (\ref{i5vc2}), with the use of (I.57), we obtain:
\begin{eqnarray}\label{i6vc1}
 I_{6}^{\mu}&=&\sum_{i=1}^{5} \, q_i^{\mu} I_{6,i},
\\\label{i6vc2}
I_{6,i}&=&
-I_{6,i}^{[d+]}
\nl
&=& (d-5) \frac{{\begin{pmatrix} 0 \\ i \end{pmatrix} }_6}{{\begin{pmatrix} 0 \\ 0 \end{pmatrix} }_6} I_{6}^{[d+]}-
\frac{1}{{\begin{pmatrix} 0 \\ 0 \end{pmatrix} }_6} \sum_{r=1}^{6} {\begin{pmatrix} 0&i \\0&r \end{pmatrix} }_6 I_{5}^{r}.
\label{second}
\end{eqnarray}
While in (\ref{i5vc2}) the first part vanishes in the limit $d \rightarrow 4$,
here its disappearance is due to (I.61):
\begin{eqnarray}
\sum_{i=1}^{5} \, q_i^{\mu} {\begin{pmatrix} 0 \\ i \end{pmatrix} }_6 =0.
\label{sumzer}
\end{eqnarray}
Indeed (\ref{sumzer}) will play a crucial role for the higher tensor
reduction. The resulting form in (\ref{i6vc2})
is already the generic form for the higher tensors too! Therefore
it appears useful to introduce the vector, applying further (A.15)
of \cite{Melrose:1965kb} and (I.61):
\begin{eqnarray}
v_r^{\mu}
&=&
-~\frac{1}{{\begin{pmatrix} 0 \\ 0 \end{pmatrix} }_6} \sum_{i=1}^{5} {\begin{pmatrix} 0&i \\  0&r\end{pmatrix}}_6 q_i^{\mu}
\nl         
&=&
-~\frac{1}{{\begin{pmatrix} 0 \\ k \end{pmatrix} }_6} \sum_{i=1}^{5} {\begin{pmatrix}0&r \\ k&i \end{pmatrix}}_6 q_i^{\mu},~~ k=0, \dots ,6 ,
\label{vr}
\end{eqnarray}
summing over all 5 (dependent) vectors. $v_r$ projected on these vectors reads:
\begin{eqnarray}
v_r \cdot q_i &=& -\frac{1}{2}\left[{\delta}_{ir}-(Y_{i6}-Y_{66})\frac{{\begin{pmatrix} 0 \\r \end{pmatrix} }_6}{{\begin{pmatrix} 0 \\ 0 \end{pmatrix} }_6}\right]
\nl              
&=& -~\frac{1}{2}\left[{\delta}_{ir}+(q_i^2+m_6^2-m_i^2) \frac{{\begin{pmatrix} 0 \\ r \end{pmatrix}}_6}{{\begin{pmatrix} 0 \\ 0 \end{pmatrix} }_6} 
\right]. 
\end{eqnarray}
 With this definition we can write in a compact way:
\begin{eqnarray}
I_{6}^{\mu}= \sum_{r=1}^{6} v_r^{\mu} E^{~r}.
\end{eqnarray}
\subsection{$R=2$ tensor integrals}
The equation (\ref{vectorred}) reads in this case:
\begin{eqnarray}
I_{6}^{\mu\, \nu} =  \sum_{i,j=1}^{5} \, q_i^{\mu}\, q_j^{\nu} \, {\nu}_{ij} \,  \, I_{6,ij}^{[d+]^2}
-\frac{1}{2}   \, g^{\mu \nu}  \, I_{6}^{[d+]} ,
\label{tensor6}
\end{eqnarray}
and by using (I.59) we have:
\begin{eqnarray}
{\nu}_{ij} \,  \, I_{6,ij}^{[d+]^2}=-(d-4) \frac{{\begin{pmatrix} 0 \\ i \end{pmatrix} }_6}{{\begin{pmatrix} 0 \\ 0 \end{pmatrix} }_6} I_{6,i}^{[d+]^2}
+\frac{{\begin{pmatrix} 0&i \\ 0&j \end{pmatrix} }_6}{{\begin{pmatrix} 0 \\ 0 \end{pmatrix} }_6} I_{6}^{[d+]}+\frac{1}{{\begin{pmatrix} 0 \\ 0 \end{pmatrix} }_6}
\sum_{r=1,r \ne i}^{6} {\begin{pmatrix}0&j \\  0&r \end{pmatrix}}_6 I_{5,i}^{[d+],r}.
\end{eqnarray}
We consider the limit $d \rightarrow 4$ and use (I.67):
\begin{eqnarray}
g^{\mu \nu}=\frac{2}{{\begin{pmatrix} 0 \\ 0 \end{pmatrix} }_6} \sum_{i,j=1}^{5} \, {\begin{pmatrix} 0&i \\ 0&j \end{pmatrix} }_6 q_i^{\mu}\, q_j^{\nu}.
\label{gmunu0}
\end{eqnarray}
Writing it like in  (\ref{tensor2}),
\be
I_{6}^{\mu\, \nu}= \sum_{i,j=1}^{5} \, q_i^{\mu}\, q_j^{\nu}  I_{6,ij},
\label{tensor6d2}
\ee
we obtain by using (\ref{first}):
\begin{eqnarray}
I_{6,ij}=-\frac{1}{{\begin{pmatrix} 0 \\ 0 \end{pmatrix} }_6}\sum_{r=1,r \ne i}^{6} {\begin{pmatrix}0&j \\ 0&r \end{pmatrix} }_6 E_i^{~r},
\end{eqnarray}
to be compared with (\ref{second}).
For completeness we specify $E_i^{~r}$, which we read off from (\ref{first}) to be:
\begin{eqnarray}
E_i^{~r}=-
\frac{1}{{\begin{pmatrix}0&r \\0&r \end{pmatrix} }_6} \sum_{s=1}^{6} {\begin{pmatrix} 0&i&r\\0&s&r \end{pmatrix} }_6 I_{4}^{rs},
\label{last}
\end{eqnarray}
and finally:
\begin{eqnarray}
I_{6}^{\mu\, \nu}=\sum_{i=1}^{5} q_i^{\mu} \sum_{r=1, r\ne i}^{6} v_r^{\nu} E_i^{~r}.
\end{eqnarray}
We remark that due to (\ref{last}), $E_i^{~r}=0$ for $r=i$ and correspondingly this will
be the case for all higher tensors such that limitations like $r \ne i$ could be dropped
but are convenient to keep in numerical programs.
\subsection{$R=3$ tensor integrals}
Equation (\ref{intone}) reads in this case:
\begin{eqnarray}
I_{6}^{\mu\, \nu\, \lambda}
  = - \sum_{i,j,k=1}^{5} \, q_i^{\mu}\, q_j^{\nu}\, q_k^{\lambda} \, {\nu}_{ij} {\nu}_{ijk} \,  \,
I_{6,ijk}^{[d+]^3}+\frac{1}{2} \sum_{i=1}^{5}
(\, g^{\mu \nu}  \, q_i^{\lambda} \,+ g^{\mu \lambda}  \, q_i^{\nu} \,+
    \, g^{\nu \lambda}  \, q_i^{\mu} \, )I_{6,i}^{[d+]^2} \, ,
\end{eqnarray}
and with (I.60) we have:
\begin{eqnarray}
{\nu}_{ij} {\nu}_{ijk}\,  \, I_{6,ijk}^{[d+]^3}=&-&(d-3) \frac{{\begin{pmatrix} 0 \\ k \end{pmatrix} }_6}{{\begin{pmatrix} 0 \\ 0 \end{pmatrix} }_6} I_{6,ij}^{[d+]^2}
+\frac{{\begin{pmatrix} 0&k\\ 0&i\end{pmatrix} }_6}{{\begin{pmatrix} 0 \\ 0 \end{pmatrix} }_6} I_{6,j}^{[d+]^2}+\frac{{\begin{pmatrix} 0&k\\ 0&j \end{pmatrix} }_6}{{\begin{pmatrix} 0 \\ 0 \end{pmatrix} }_6} I_{6,i}^{[d+]^2}\nn \\
&+&\frac{1}{{\begin{pmatrix} 0 \\ 0 \end{pmatrix} }_6} \sum_{r=1,r \ne i,j}^{6} {\begin{pmatrix} 0&k\\ 0&r\end{pmatrix} }_6 {\nu}_{ij} I_{5,ij}^{[d+]^2,r}.
\end{eqnarray}
The first term on the r.h.s. is eliminated due to (\ref{sumzer}) and the next two terms cancel
due to (\ref{gmunu0}).
Taking into account $I_5^{[d+]}$, relation (\ref{tenstwo}) now reads:
\begin{eqnarray}
I_{5,ij}^r={\nu}_{ij} I_{5,ij}^{[d+]^2,r}-\frac{{\begin{pmatrix}i&r \\ j&r\end{pmatrix} }_6}{{\begin{pmatrix} r\\ r\end{pmatrix} }_6}I_{5}^{[d+],r}.
\end{eqnarray}
As a further representation of $g^{\mu \nu}$ we have (see (I.75)):
\begin{eqnarray}
g^{\mu \nu}=\frac{2}{{\begin{pmatrix}r \\  r\end{pmatrix}}_6} \sum_{i,j=1}^{5} \, {\begin{pmatrix} i&r\\  j&r\end{pmatrix}}_6 q_i^{\mu}\, q_j^{\nu},~~~
r=1 \dots 6~.
\label{gmunur}
\end{eqnarray}
Using again (I.57) and the definition
\be
I_{6}^{\mu\, \nu \lambda}= \sum_{i,j,k=1}^{5} \, q_i^{\mu}\, q_j^{\nu} q_k^{\lambda} I_{6,ijk},
\label{tensor6d3}
\ee
we obtain:
\begin{eqnarray}
I_{6,ijk}=-\frac{1}{{\begin{pmatrix} 0 \\ 0 \end{pmatrix} }_6}\sum_{r=1,r \ne i,j}^{6} {\begin{pmatrix} 0&k\\ 0&r\end{pmatrix} }_6 I_{5,ij}^{r}.
\end{eqnarray}
From (\ref{final2}) and (\ref{gmunur}) $I_{5,ij}^{r}$ reads:
\begin{eqnarray}
 I_{5,ij}^{r}= E_{ij}^{~r} + 2 \frac{{\begin{pmatrix} i&r\\  j&r\end{pmatrix}}_6}{{\begin{pmatrix} r\\r \end{pmatrix} }_6} E_{00}^r ,
\end{eqnarray}
so that we get:
\begin{eqnarray}
I_{6}^{\mu\, \nu\, \lambda}= \sum_{i,j=1}^{5} q_i^{\mu} q_j^{\nu}  \sum_{r=1,r \ne i,j}^{6}
                            v_r^{\lambda}  E_{ij}^{~r}
                            + \sum_{i,j=1}^{5} q_i^{\mu} q_j^{\nu}  \sum_{r=1}^{6}
2\frac{{\begin{pmatrix}i&r \\ j&r \end{pmatrix}}_6 }{{\begin{pmatrix} r\\r \end{pmatrix} }_6}        v_r^{\lambda}  E_{00}^{~r},
\label{after}
\end{eqnarray}
where in the second term we can drop the limitation $r \ne i,j$ since it is automatically
fulfilled due to the numerator $ {\begin{pmatrix}i&r \\ j&r\end{pmatrix} }_6$, vanishing for $r=i$ and $r=j$. Thus summation
over $i$ and $j$ is possible, using (\ref{gmunur}), with a result:
\begin{eqnarray}
I_{6}^{\mu\, \nu\, \lambda}= \sum_{i,j=1}^{5} q_i^{\mu} q_j^{\nu}  \sum_{r=1,r \ne i,j}^{6}
                            v_r^{\lambda}  E_{ij}^{~r}
                            + g^{\mu \nu}  \sum_{r=1}^{6} v_r^{\lambda}  E_{00}^{~r} ,
\end{eqnarray}
or:
\begin{eqnarray}
I_{6}^{\mu\, \nu\, \lambda}
&=&\sum_{r=1}^{6}  v_r^{\lambda} I_{5}^{\mu\, \nu\, ,r},
\end{eqnarray}
with
\begin{eqnarray}
I_{5}^{\mu\, \nu\, ,r}&=& 
\sum_{i,j=1, i,j\ne r}^{5} q_i^{\mu} q_j^{\nu}  E_{ij}^{~r}+ g^{\mu \nu}  E_{00}^{~r} .
\end{eqnarray}
\subsection{$R=4$ tensor integrals}
The tensor integral in (\ref{tensorred1}) contains three different integrals in
higher dimension, which have to be reduced or to be eliminated. We begin with
$I_{n,ijkl}^{[d+]^4}$ using (I.26). For convenience we use $x$ instead of $4$:
\begin{eqnarray}
{\begin{pmatrix} 0 \\ 0 \end{pmatrix} }_n {\nu}_{ijkl}~ {\bf l^+}~ I_{n,ijk }^{[d+]^x} 
&\equiv&
{\begin{pmatrix} 0 \\ 0 \end{pmatrix} }_n {\nu}_{ijkl} I_{n,ijkl}^{[d+]^x}
\nl
&=& \sum_{r=1}^n {\begin{pmatrix}0&l \\ 0&r \end{pmatrix}}_n \left[d+2 x -(n+3) \right]I_{n,ijk }^{[d+]^x} 
\nn \\
&&-~\sum_{s=1}^n {\begin{pmatrix}0&l \\ 0&s \end{pmatrix} }_n  {\nu}_{ijks} I_{n,ijk }^{[d+]^x}-
\sum_{r,s=1; r \neq s}^n {\begin{pmatrix} 0&l\\ 0&r\end{pmatrix} }_n {\nu}_{ijks}~{\bf r^-}~{\bf s^+}~I_{n,ijk }^{[d+]^x} 
\nn\\
&=&\left\{\left[n+4-(d+2 x)\right] {\begin{pmatrix} 0 \\ l\end{pmatrix} }_n- {\begin{pmatrix}0&l \\0&i \end{pmatrix} }_n-{\begin{pmatrix}0&l \\ 0&j \end{pmatrix} }_n-{\begin{pmatrix} 0&l\\ 0&k \end{pmatrix} }_n \right \} I_{n,ijk }^{[d+]^x}
\nn\\
&&-~\sum_{r=1}^n {\begin{pmatrix} 0&l\\ 0&r\end{pmatrix} }_n\sum_{s=1;s \neq r}^n {\nu}_{ijks} I_{n,ijks}^{[d+]^x,r}.
\label{I4}
\end{eqnarray}
The last double sum in (\ref{I4}), assuming all indices $i,j,k$ to be different,
reads:
\begin{eqnarray}
&&-{\begin{pmatrix}0&l \\  0&i\end{pmatrix}}_n \sum_{s=1;s \neq i }^n  {\nu}_{jks} I_{n,jks }^{[d+]^x}
-{\begin{pmatrix} 0&l\\ 0&j \end{pmatrix} }_n \sum_{s=1;s \neq j }^n  {\nu}_{iks} I_{n,iks }^{[d+]^x}
-{\begin{pmatrix}0&l \\ 0&k \end{pmatrix} }_n \sum_{s=1;s \neq k }^n  {\nu}_{ijs} I_{n,ijs }^{[d+]^x} \nn\\
&&- \sum_{r=1;r \neq i,j,k}^n {\begin{pmatrix} 0&l\\0&r \end{pmatrix} }_n \sum_{s=1;s \neq r}^n {\nu}_{ijks} I_{n-1,ijks}^{[d+]^x,r}.
\label{ijk}
\end{eqnarray}
Now adding corresponding terms in (\ref{I4}) and (\ref{ijk}), e.g. for $r=i$,
we get:
\begin{eqnarray}
-{\begin{pmatrix} 0&l \\ 0&i \end{pmatrix}}_n \sum_{s=1;s \neq i }^n  {\nu}_{jks} I_{n,jks }^{[d+]^x}-
 {\begin{pmatrix} 0&l \\ 0&i \end{pmatrix}}_n I_{n,ijk }^{[d+]^x}=
-{\begin{pmatrix} 0&l \\ 0&i \end{pmatrix}}_n \sum_{s=1}^n  {\nu}_{jks} I_{n,jks }^{[d+]^x}={\begin{pmatrix} 0&l \\ 0&i \end{pmatrix}}_n
I_{n,jk }^{[d+]^{(x-1)}},  \nn\\
\label{Now}
\end{eqnarray}
due to (I.29). In case two indices are equal, e.g. $i=j \ne k$, we have:
\begin{eqnarray}
-{\begin{pmatrix} 0&l \\ 0&i \end{pmatrix}}_n \sum_{s=1;s \neq i }^n  (1+2 \delta_{is}+\delta_{ks}) I_{n,iks }^{[d+]^x}-2 {\begin{pmatrix} 0&l \\ 0&i \end{pmatrix}}_n I_{n,iis }^{[d+]^x}
&=&-{\begin{pmatrix} 0&l \\ 0&i \end{pmatrix}}_n \sum_{s=1 }^n
(1+ \delta_{is}+\delta_{ks})I_{n,iks }^{[d+]^x} \nn\\
&\equiv& -{\begin{pmatrix} 0&l \\ 0&i \end{pmatrix}}_n \sum_{s=1 }^n {\nu}_{iks} I_{n,iks }^{[d+]^x}, \nn\\
\end{eqnarray}
like (\ref{Now}), i.e. if two indices agree, this integral occurs only once. 
As final result
we have:
\begin{eqnarray}
{\begin{pmatrix} 0 \\ 0 \end{pmatrix} }_n {\nu}_{ijkl}~ I_{n,ijkl}^{[d+]^x} &=&
\left[n+4-(d+2 x)\right]
{\begin{pmatrix} 0 \\ l\end{pmatrix} }_n  I_{n,ijk}^{[d+]^x} \nn \\
&+&\left[ijk\right]^{(l)}_{\mathrm red} +  \sum_{r=1;r \ne i,j,k }^n 
{\begin{pmatrix} 0&l \\  0&r\end{pmatrix} }_n
 I_{n-1,ijk}^{[d+]^{(x-1)},r},
\label{final}
\end{eqnarray}
where according to (\ref{Now}) and the discussion thereafter:
\begin{eqnarray}
\left[ijk\right]^{(l)}={\begin{pmatrix} 0&l \\ 0&i \end{pmatrix}}_n I_{n,jk}^{[d+]^{(x-1)}}+
                       {\begin{pmatrix} 0&l \\ 0&j \end{pmatrix} }_n I_{n,ik}^{[d+]^{(x-1)}}+
                       {\begin{pmatrix} 0&l \\ 0&k \end{pmatrix} }_n I_{n,ij}^{[d+]^{(x-1)}}
\end{eqnarray}
and $\left[ijk\right]^{(l)}_{\mathrm red}=\left[ijk\right]^{(l)}$ without repetition, e.g.
$\left[iii\right]^{(l)}_{\mathrm red}={\begin{pmatrix} 0&l \\ 0&i \end{pmatrix}}_n I_{n,ii}^{[d+]^{(x-1)}}$.

\vspace{1cm}

Now, making use of $n_{ijkl}={\nu}_{ij}{\nu}_{ijk}{\nu}_{ijkl}$, we see that
due to (\ref{sumzer}) the first part in (\ref{final}) drops out after insertion into
(\ref{tensorred1}). The second contribution of (\ref{final}) yields:
\begin{eqnarray}\label{432}
\frac{1}{{\begin{pmatrix} 0 \\ 0 \end{pmatrix} }_n} \sum_{l=1}^{n-1} q_l^{\rho} \sum_{i,j,k=1}^{n-1}{\nu}_{ij}{\nu}_{ijk} \left[ijk\right]^{(l)}_{\mathrm red}  q_i^{\mu} q_j^{\nu} q_k^{\lambda}.
\end{eqnarray}
We have:
\begin{eqnarray}
{\nu}_{ij}{\nu}_{ijk}\left[ijk\right]^{(l)}_{\mathrm red}
&=&
\left[ijk\right]^{(l)}
+
{\begin{pmatrix} 0&l \\ 0&i \end{pmatrix}}_n \delta_{jk} I_{n,jk}^{[d+]^{(x-1)}}
\nl &&
+
{\begin{pmatrix} 0&l \\ 0&j \end{pmatrix} }_n \delta_{ik} I_{n,ik}^{[d+]^{(x-1)}}
+
{\begin{pmatrix} 0&l \\ 0&k \end{pmatrix} }_n \delta_{ij} I_{n,ij}^{[d+]^{(x-1)}} ,
\end{eqnarray}
with the help of which (\ref{432}) reads:
\begin{eqnarray}
\frac{1}{{\begin{pmatrix} 0 \\ 0 \end{pmatrix} }_n} \sum_{l=1}^{n-1} q_l^{\rho} \sum_{i,j,k=1}^{n-1}
\left[ q_i^{\mu} {\begin{pmatrix} 0&l \\ 0&i \end{pmatrix}}_n (1+\delta_{jk}) I_{n,jk}^{[d+]^{(x-1)}}  q_j^{\nu} q_k^{\lambda}+
       q_j^{\nu} {\begin{pmatrix} 0&l \\ 0&j \end{pmatrix} }_n (1+\delta_{ik}) I_{n,ik}^{[d+]^{(x-1)}}  q_i^{\mu} q_k^{\lambda}
       \right. \nn\\ \left.
 + q_k^{\lambda} {\begin{pmatrix} 0&l \\ 0&k \end{pmatrix} }_n (1+\delta_{ij}) I_{n,ij}^{[d+]^{(x-1)}}  q_i^{\mu} q_j^{\nu} \right].
\nl
\end{eqnarray}
Using (I.67) we have for $d=4$:
\begin{eqnarray}
\frac{1}{2}\left\{g^{\mu \rho} n_{jk} I_{n,jk}^{[d+]^{(x-1)}} q_j^{\nu} q_k^{\lambda} +
                  g^{\nu \rho} n_{ik} I_{n,ik}^{[d+]^{(x-1)}} q_i^{\mu} q_k^{\lambda} +
              g^{\lambda \rho} n_{ij} I_{n,ij}^{[d+]^{(x-1)}} q_i^{\mu} q_j^{\nu} \right\} ,
\end{eqnarray}
and we see that this contribution is canceled by the last three terms
of the type $I_{n,jk}^{[d+]^{(x-1)}}$ in (\ref{tensorred1}).
The first three terms of this
type are evaluated by means of (I.59) to yield:
\begin{eqnarray}
n_{ij} I_{n,jk}^{[d+]^{(x-1)}}
&=&
\frac{1}{{\begin{pmatrix} 0 \\ 0 \end{pmatrix} }_n}
       \left\{  \left[n+2(2-x)-d \right]{\begin{pmatrix} 0 \\ j \end{pmatrix} }_n  I_{n,i}^{[d+]^{(x-1)}}
+                 {\begin{pmatrix} 0&i \\ 0&j \end{pmatrix} }_n I_{n}^{[d+]^{(x-2)}}    \right. \nn\\
&&
\left. +~\sum_{r=1;r \neq i }^n {\begin{pmatrix}0&j \\  0&r \end{pmatrix}}_n I_{n-1,i}^{[d+]^{(x-2)},r} \right\}.
\end{eqnarray}
Inserting this into (\ref{tensorred1}), the first part yields a vanishing
contribution due to (\ref{sumzer}) .
The second term yields, again due to (\ref{sumzer}):
\begin{eqnarray}
&&-\frac{1}{2 {\begin{pmatrix} 0 \\ 0 \end{pmatrix} }_n} \sum_{i,j=1}^{n-1}
\left\{g^{\mu \nu}  q_i^{\lambda} q_j^{\rho} +
             g^{\mu \lambda} q_i^{\nu} q_j^{\rho} +
             g^{\nu \lambda} q_i^{\mu} q_j^{\rho} \right\}  {\begin{pmatrix} 0&i \\ 0&j \end{pmatrix} }_n I_{n}^{[d+]^{(x-2)}}\nn\\
&=&-\frac{1}{4} \left(g^{\mu \nu} g^{\lambda \rho}+g^{\mu \lambda} g^{\nu \rho}
                     +g^{\mu \rho} g^{\nu \lambda}\right)I_{n}^{[d+]^{(x-2)}},
\end{eqnarray}
which cancels the last term in  (\ref{tensorred1})
and the total contribution thus reads:
\begin{eqnarray}
\frac{1}{ {\begin{pmatrix} 0 \\ 0 \end{pmatrix} }_n} \left\{ \sum_{i,j,k,l=1}^{n-1}  {\nu}_{ij}{\nu}_{ijk}
      \, q_i^{\mu}\, q_j^{\nu}\, q_k^{\lambda}  \, q_l^{\rho}\,
 \sum_{r=1;r  \ne i,j,k }^{n}
 {\begin{pmatrix} 0&l \\ 0&r \end{pmatrix}}_n
 I_{n-1,ijk}^{[d+]^{(x-1)},r} \right. \nn\\
\left. -\frac{1}{2} \sum_{i,j=1}^{n-1} \left(g^{\mu \nu}  q_i^{\lambda} q_j^{\rho} +
             g^{\mu \lambda} q_i^{\nu} q_j^{\rho} +
             g^{\nu \lambda} q_i^{\mu} q_j^{\rho} \right) \sum_{r=1;r  \ne i }^{n}
             {\begin{pmatrix}0&j \\   0&r\end{pmatrix}}_n I_{n-1,i}^{[d+]^{(x-2)},r} \right\},
\label{pcancel}
\end{eqnarray}
reducing the 6-point tensor to 5-point tensors in lower dimensions. For further
reduction we put explicitly $n=6$ and $x=4$ and write  (\ref{raw3}) in the form
\begin{eqnarray}
{\nu}_{ij} {\nu}_{ijk} I_{5,ijk}^{[d+]^3,r} =-I_{5,ijk}^r+
\left[\frac{{\begin{pmatrix} j&r\\  k&r\end{pmatrix}}_6}{{\begin{pmatrix}r \\ r\end{pmatrix} }_6} I_{5,i}^{[d+]^2,r}+
      \frac{{\begin{pmatrix}i&r \\k&r \end{pmatrix} }_6}{{\begin{pmatrix} r\\r \end{pmatrix} }_6} I_{5,j}^{[d+]^2,r}+
      \frac{{\begin{pmatrix} i&r\\j&r \end{pmatrix} }_6}{{\begin{pmatrix}r \\ r\end{pmatrix} }_6} I_{5,k}^{[d+]^2,r} \right].
\label{rawinv}
\end{eqnarray}
With  (\ref{gmunur}) it is now easy to see that the square bracket in  (\ref{rawinv})
cancels out the second part in (\ref{pcancel})
and using the definition:
\be
I_{6}^{\mu\, \nu \, \lambda \, \rho}
 = \sum_{i,j,k,l=1}^{5} \, q_i^{\mu}\, q_j^{\nu} q_k^{\lambda} q_l^{\rho}
 I_{6,ijkl},
\label{tensor6d4}
\ee
we obtain:
\begin{eqnarray}
I_{6,ijkl}=-\frac{1}{{\begin{pmatrix} 0 \\ 0 \end{pmatrix} }_6}
\sum_{r=1,r \ne i,j,k}^{6} {\begin{pmatrix} 0&l\\ 0&r\end{pmatrix} }_6
 I_{5,ijk}^{r}.
\end{eqnarray}
Again, with (\ref{final3}) and (\ref{gmunur}) $I_{5,ijk}^{r}$ reads:
\begin{eqnarray}
I_{5,ijk}^{r}= E_{ijk}^{~r}
 + 2 \frac{{\begin{pmatrix}i&r \\j&r \end{pmatrix} }_6}
          {{\begin{pmatrix}r \\r \end{pmatrix} }_6} E_{00k}^r
 + 2 \frac{{\begin{pmatrix} i&r\\k&r \end{pmatrix} }_6}
          {{\begin{pmatrix} r\\r \end{pmatrix} }_6} E_{00j}^r
 + 2 \frac{{\begin{pmatrix} j&r\\k&r \end{pmatrix} }_6}
          {{\begin{pmatrix} r\\r \end{pmatrix} }_6} E_{00i}^r ,
\end{eqnarray}
so:
\begin{eqnarray}
I_{6}^{\mu\, \nu\, \lambda \rho}=
&& \sum_{i,j,k=1}^{5} q_i^{\mu} q_j^{\mu} q_k^{\lambda}
\sum_{r=1,r \ne i,j,k}^{6} v_r^{\rho}  E_{ijk}^{~r}
\nn \\
&& {}+\sum_{i,j,k=1}^{5} q_i^{\mu} q_j^{\mu}  q_k^{\lambda}
\sum_{r=1,r \ne i,j,k}^{6}  v_r^{\rho} \left\{
   2 \frac{{\begin{pmatrix}i&r \\ j&r\end{pmatrix} }_6}
          {{\begin{pmatrix}r \\r \end{pmatrix} }_6} E_{00k}^r
   + 2 \frac{{\begin{pmatrix} i&r\\ k&r\end{pmatrix} }_6}
            {{\begin{pmatrix} r\\r \end{pmatrix} }_6} E_{00j}^r
   + 2 \frac{{\begin{pmatrix} j&r\\k&r \end{pmatrix} }_6}
            {{\begin{pmatrix} r\\r \end{pmatrix} }_6} E_{00i}^r \right\} ,
\nl
\end{eqnarray}
and with the same argument like the one used after (\ref{after}) we
obtain the final result:
\begin{eqnarray}
I_{6}^{\mu\, \nu\, \lambda \rho}&=&
 \sum_{i,j,k=1}^{5} q_i^{\mu} q_j^{\mu} q_k^{\lambda}
\sum_{r=1,r \ne i,j,k}^{6} v_r^{\rho}  E_{ijk}^{~r} 
+ 
 g^{\mu \nu}     \sum_{k=1}^{5} q_k^{\lambda}  \sum_{r=1,r \ne k}^{6} v_r^{\rho} E_{00k}^r
\nn \\
&&
+~g^{\mu \lambda} \sum_{j=1}^{5} q_j^{\nu}      \sum_{r=1,r \ne j}^{6} v_r^{\rho} E_{00j}^r
+g^{\nu \lambda} \sum_{i=1}^{5} q_i^{\mu}      \sum_{r=1,r \ne i}^{6} v_r^{\rho} E_{00i}^r ,
\end{eqnarray}
or:
\begin{eqnarray}
I_{6}^{\mu\, \nu\, \lambda \rho}
&=&
\sum_{r=1}^{6}  v_r^{\rho} I_{5}^{\mu\, \nu\, \lambda ,r} ,
\end{eqnarray}
with:
\begin{eqnarray}
I_{5}^{\mu\, \nu\, \lambda ,r}
&=& \sum_{i,j,k=1; i,j,k \ne r}^{5} q_i^{\mu} q_j^{\nu} q_k^{\lambda}
E_{ijk}^{~r}+  \sum_{k=1, k \ne r}^5 g^{[\mu \nu} q_k^{\lambda]} E_{00k}^{~r}.
\end{eqnarray}

\section{\label{Numerical}Numerical results and discussion}
In order to illustrate the numerical results which can be obtained with
the described approach, we will evaluate a representative collection of
tensor coefficients.
We rely on two implementations of the formalism, one has been
established in Fortran, and the other one in the Mathematica package
\texttt{hexagon.m}.

\begin{figure}[t]
\begin{center}
\includegraphics[scale=0.9]{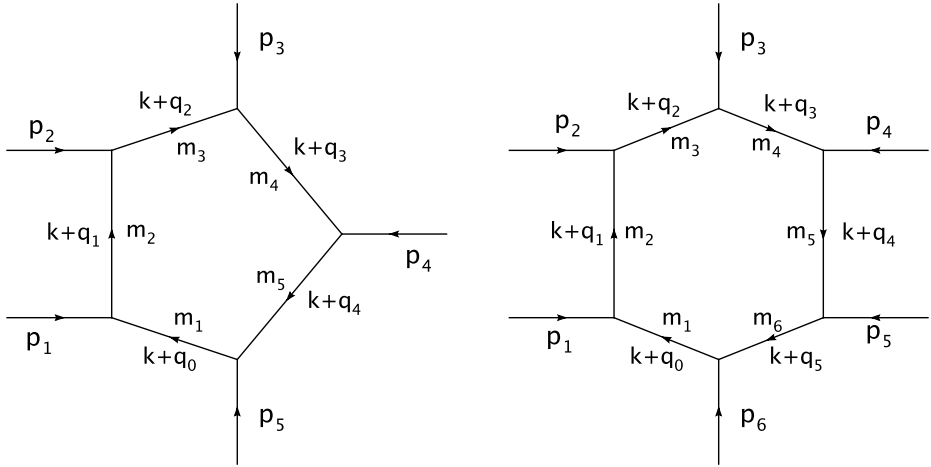}
\caption{
Momenta flow used in the numerical examples for six- and five-point integrals.}
\label{fig:6pt5ptfig}
\end{center}
\end{figure}

In the following, we denote the scalar five-point function by $E_0$ and the scalar six-point function  by $F_0$.
The tensor decompositions of pentagons $E$ and hexagons $F$ read:
\begin{eqnarray}
	E^\mu &=& \sum_{i=1}^{4}q_i^\mu E_i,
\\
	E^{\mu \nu} &=& \sum_{i,j=1}^{4}q_i^\mu q_j^\nu E_{ij}+
		g^{\mu \nu} E_{00},
\\
	E^{\mu \nu \lambda} &=& 
		\sum_{i,j,k=1}^{4}q_i^\mu q_j^\nu q_k^\lambda E_{ijk}+
		\sum_{i=1}^{4} 
		g^{[\mu \nu} q_i^{\lambda ]} E_{00i},
\\
	F^\mu &=& \sum_{i=1}^{5}q_i^\mu F_i,
\\
	F^{\mu \nu} &=& \sum_{i,j=1}^{5}q_i^\mu q_j^\nu F_{ij},
\\
   F^{\mu \nu \lambda} &=& 
		\sum_{i,j,k=1}^{5}q_i^\mu q_j^\nu q_k^\lambda F_{ijk}+
		\sum_{i=1}^{5} 
		g^{\mu \nu} q_i^{\lambda} F_{00i},	
\\
	F^{\mu \nu \lambda \rho} &=&		
	\sum_{i,j,k,l=1}^{5}q_i^{\mu} q_j^{\nu} q_k^{\lambda} q_l^{\rho} F_{ijkl}
	+
		\sum_{i,j=1}^{5} 
		q_i^{\mu} q_j^{[ \nu} g^{\lambda \rho ] } F_{00ij}.
\end{eqnarray}
Please observe the difference of $E^0,F^0$ and $E_0,F_0$ in the following. 
The kinematics is visualized in Figure ~\ref{fig:6pt5ptfig}.
Deviating from the first sections, we have chosen here $q_0=0$ in order to stay close to common conventions of other numerical packages.

For the evaluation of the scalar two-, three- and four-point functions,
which appear after the complete reduction,
we have implemented two numerical libraries:
\begin{itemize}
\item 
For massive internal particles:
\texttt{Looptools 2.2} \cite{Hahn:1998yk,Hahn:2006qw}; 
\item 
If there are also massless internal particles:
\texttt{QCDLoop-1.4} \cite{Ellis:2007qk}.
\end{itemize}
We observed that \texttt{Looptools} may become unstable in the presence of massless internal particles, while \texttt{QCDLoop} seems to be generally slower.
Our Mathematica package has an implementation of only \texttt{Looptools}.

For completeness, we would like to mention also other publicly available Fortran packages for tensor functions, which we found useful for comparisons:
\begin{itemize}
\item Six-point tensors with massive internal particles: none;
\item Five-point tensors with massive internal particles:
 \texttt{Looptools} \cite{Hahn:1998yk,Hahn:2006qw} ;
\item Five-point tensors with both massive and massless particles: none; 
\item Five- and six-point tensors with only massless internal particles:
\texttt{golem95} \cite{Binoth:2008uq}.
\end{itemize}

The two independent numerical implementations have been checked in several ways:
\begin{itemize}
 \item
By internal comparisons of the two codes, relying on the formulae presented in this article;
\\
With alternative, direct representations of the tensor integrals with
sector decomposition
\footnote{%
We used a  Mathematica 
interface to the \texttt{GINAC} package \texttt{sector\_decomposition}
in order to have a convenient way to evaluate tensor Feynman integrals.
}
 \cite{Bogner:2007cr} and Mellin-Barnes representations
\cite{Gluza:2007rt,Czakon:2005rk};
\item
By simplifying the numerator structures algebraically and subsequent
evaluation of the resulting integrals of lower rank;
\item
By direct comparison with other tensor integral packages \cite{Hahn:2006qw,Binoth:2008uq}.
\end{itemize}
Some of the comparisons were documented in \cite{Diakonidis:2008dt}.
 
We restrict ourselves to a few phase-space points,
see Tables \ref{phasespacepoint} to \ref{phasespacepoint-1}.
The first configuration corresponds to the reaction
$gg\to t{\bar t}q{\bar q}$, with external momenta generated by
\texttt{Madgraph}~\cite{Stelzer:1994ta,Maltoni:2002qb}.
The second configuration comes from \cite{Binoth:2008uq}, while the third
is a slight modification of the first one.
The kinematical input is completed by adding the masses of internal particles.

\begin{table}[b]
\centering
\begin{tabular}{|r|r|r@{.}l|r@{.}l|r|}
\hline
$p_1$ & 0.21774554~E+03   & 0&0 & 0&0 & 0.21774554~E+03 \\ 
$p_2$ & 0.21774554~E+03   & 0&0 & 0&0 &  -- 0.21774554~E+03\\ 
$p_3$ & -- 0.20369415~E+03 &-- 0&47579512~E+02   &0&42126823~E+02   &0.84097181~E+02   \\ 
$p_4$ & -- 0.20907237~E+03 & 0&55215961~E+02  & -- 0&46692034~E+02  &  -- 0.90010087~E+02 \\ 
$p_5$ &  -- 0.68463308~E+01& 0&53063195~E+01  &  0&29698267~E+01  & -- 0.31456871~E+01  \\ 
$p_6$ & -- 0.15878244~E+02 &  -- 0&12942769~E+02 & 0&15953850~E+01  &  0.90585932~E+01
\\
\hline
\multicolumn{7}{|c|}{$m_1 = 110.0, \;
m_2 = 120.0, ~~\;
m_3 = 130.0, ~~\;
m_4 = 140.0, ~~\;
m_5 = 150.0, ~~\;
m_6 = 160.0$}
\\
\hline
\end{tabular}
\caption{\label{phasespacepoint}
The components of external four-momenta for the six-point numerics;
all internal particles are massive. For five-point functions, we shrink
line 2 and fix $p_1+p_2 \to p_1$ in order to retain momentum conservation.
}
\end{table}

\begin{table}[tb]
\centering
\begin{tabular}{|r|r@{.}l|r@{.}l|r@{.}l|r@{.}l|}
\hline
$p_1$ & 0&5 & 0&0 & 0&0 & 0&5 \\ 
$p_2$ & 0&5 & 0&0 & 0&0 &  -- 0&5\\ 
$p_3$ &  -- 0&19178191 & -- 0&12741180 &-- 0&08262477 &-- 0&11713105  \\ 
$p_4$ & -- 0&33662712 & 0&06648281& 0&31893785& 0&08471424\\ 
$p_5$ & -- 0&21604814& 0&20363139&-- 0&04415762& -- 0&05710657   
\\
\hline
\multicolumn{9}{|c|}{$p_6=-(p_1+p_2+p_3+p_4+p_5)$, ~~~~$m_1 = \cdots = m_6 = 0.0$}
\\
\hline
\end{tabular}
\caption{\label{massless-phasespacepoint}
The external four-momenta  for the six-point numerics;
all internal particles are massless. This set of momenta
comes from \cite{Binoth:2008uq}.
For five-point functions, we shrink line 2 and fix $p_1+p_2 \to p_1$
in order to retain momentum conservation.}
\end{table}

\begin{table}[b]
\centering
\begin{tabular}{|r|r@{.}l|r@{.}l|r@{.}l|r@{.}l|}
\hline
$p_1$ & 0&21774554~E+01 & 0&0 & 0&0 & 0&21774554~E+01 \\
$p_2$ & 0&21774554~E+01 & 0&0 & 0&0 &  -- 0&21774554~E+01\\
$p_3$ & -- 0&20369415~E+01 &-- 0&47579512~E+00   &0&42126823~E+00   &0&84097181~E+00   \\
$p_4$ & -- 0&20907237~E+01  & 0&55215961~E+00  & -- 0&46692034~E+00  &  -- 0&90010087~E+00 \\
$p_5$ &  -- 0&68463308~E--01 & 0&53063195~E--01  &  0&29698267~E--01  & -- 0&31456871~E--01  \\
$p_6$ & -- 0&15878244~E+00  &  -- 0&12942769~E+00 & 0&15953850~E--01  &  0&90585932~E--01
\\
\hline
\multicolumn{9}{|c|}{$m_1 = 0.0, \;
m_2 = 0.0, ~~\;
m_3 = 0.0, ~~\;
m_4 = 1.7430, ~~\;
m_5 = 0.0, ~~\;
m_6 = 0.0$}
\\
\hline
\end{tabular}
\caption{\label{phasespacepoint-1}
The external four-momenta  for the six-point numerics; one internal mass
is finite.  For five-point functions, we shrink line 2 and fix
$p_1+p_2 \to p_1$ in order to retain momentum conservation.}
\end{table}

We begin with massive six-point tensors.
For the kinematics introduced above, we determine the tensor components with
our Fortran pacakge as shown in
Tables \ref{tab:hexagonResults} to \ref{tab:hexagonResults4rank}.
They are complex, finite numbers.
Only independent components of the tensors are shown, all the remaining
ones are obtained by permutations of indices.

\begin{table}[htb]
\centering
\vspace{5pt}
\begin{tabular}{|c c|r@{ i }l|}
\hline
& &\multicolumn{2}{|c|}{$F_0$} \\ \hline
& &-- 0.223393 E--18      -- & 0.396728 E--19\\\hline
$\mu$ & &\multicolumn{2}{|c|}{$F^{\mu}$} \\ \hline
0&     &0.192487 E--17    +  &   0.972635 E--17 \\ 
1&    &-- 0.363320 E--17  -- & 0.11940 E--17  \\ 
2&     &0.365514 E--17     + &   0.106928 E--17 \\ 
3&     &0.239793 E--16     + &   0.341928 E--17 \\ \hline
$\mu$&$\nu$&\multicolumn{2}{|c|}{$F^{\mu\nu}$} \\ \hline
0&0&   0.599459 E--14    --  & 0.114601 E--14 \\ 
0&1&    0.323869 E--15     + &  0.423754 E--15\\ 
0&2&   -- 0.294252 E--15 --  & 0.375481 E--15\\ 
0&3&   -- 0.255450 E--14  -- &  0.195640 E--14\\ 
1&1&   -- 0.164562 E--14  -- &  0.993796 E--16\\ 
1&2&    0.920944 E--16     + &  0.706487 E--17\\ 
1&3&    0.347694 E--15    -- & 0.127190 E--16\\ 
2&2&   -- 0.163339 E--14 --  &  0.994148 E--16\\ 
2&3&   -- 0.341773 E--15   + &  0.818678 E--17\\ 
3&3&   -- 0.413909 E--14   + &  0.670676 E--15\\ \hline
\end{tabular}
\caption[Tensor components  for scalar, vector, and rank $R=2$ six-point functions.]
{\label{tab:hexagonResults}Tensor components  for scalar, vector, and rank $R=2$ six-point functions; kinematics defined in Table \ref{phasespacepoint} and Figure \ref{fig:6pt5ptfig}. }
\end{table}

\begin{table}[htb]
\label{tab:hexagonResults3rank}
\centering
\vspace{5pt}
\begin{tabular}{|l l c|r@{ i }l|}
\hline
$\mu$ &$\nu$&$\lambda$ &\multicolumn{2}{|c|}{$F^{\mu\nu\lambda}$} \\ \hline
0&0&0&         -- 0.227754 E--11 -- &          0.267244 E--12 \\
0&0&1&          0.140271 E--13 -- &         0.119448 E--12 \\
0&0&2&         -- 0.201270 E--13  + &          0.101968 E--12 \\
0&0&3&          0.102976 E--12  + &          0.624467 E--12 \\
0&1&1&          0.183904 E--12  + &          0.142429 E--12 \\
0&1&2&         -- 0.131028 E--13 -- &          0.610343 E--14 \\
0&1&3&         -- 0.543316 E--13 -- &          0.158809 E--13 \\
0&2&2&          0.181352 E--12  + &          0.141686 E--12 \\
0&2&3&          0.506408 E--13  + &          0.163568 E--13 \\
0&3&3&          0.600542 E--12  + &          0.130733 E--12 \\
1&1&1&         -- 0.563539 E--13  + &          0.178403 E--13 \\
1&1&2&          0.210641 E--13 -- &          0.584990 E--14 \\
1&1&3&          0.120482 E--12 -- &          0.574688 E--13 \\
1&2&2&         -- 0.201182 E--13  + &          0.620591 E--14 \\
1&2&3&         -- 0.686164 E--14  + &          0.205457 E--14 \\
1&3&3&         -- 0.447329 E--13  + &          0.193180 E--13 \\
2&2&2&          0.582201 E--13 -- &          0.163889 E--13 \\
2&2&3&          0.119659 E--12 -- &          0.570084 E--13 \\
2&3&3&          0.457464 E--13 -- &          0.181141 E--13 \\
3&3&3&          0.557081 E--12 -- &          0.374359 E--12 \\ \hline
\end{tabular}
\caption[Tensor components for a massive rank $R=3$ six-point function]
{Tensor  components for a massive rank $R=3$ six-point function; kinematics defined in Table \ref{phasespacepoint} and Figure \ref{fig:6pt5ptfig}.}
\end{table}

\begin{table}[htb]
\centering
\begin{tabular}{|l l c c|r@{ i }l|}
\hline
$\mu$ &$\nu$ &$\lambda$ &$\rho$ &\multicolumn{2}{|c|}{$F^{\mu\nu\lambda\rho}$} \\ \hline
0&0&0&0&          0.666615 E--09  + &          0.247562 E--09 \\
0&0&0&1&         -- 0.200049 E--10  + &          0.294036 E--10 \\
0&0&0&2&          0.200975 E--10 -- &          0.237333 E--10 \\
0&0&0&3&          0.645477 E--10 -- &          0.162236 E--09 \\
0&0&1&1&         -- 0.116956 E--10 -- &          0.516760 E--10 \\
0&0&1&2&          0.160357 E--11  + &          0.222284 E--11 \\
0&0&1&3&          0.792692 E--11  + &          0.729502 E--11 \\
0&0&2&2&         -- 0.111838 E--10 -- &          0.513133 E--10 \\
0&0&2&3&         -- 0.681086 E--11 -- &          0.708933 E--11 \\
0&0&3&3&         -- 0.804454 E--10 -- &          0.801909 E--10 \\
0&1&1&1&          0.100498 E--10 -- &          0.151735 E--13 \\
0&1&1&2&         -- 0.348984 E--11 -- &          0.195436 E--12 \\
0&1&1&3&         -- 0.211111 E--10  + &          0.295212 E--11 \\
0&1&2&2&          0.357455 E--11  + &          0.662809 E--14 \\
0&1&2&3&          0.121595 E--11 -- &          0.807388 E--13 \\
0&1&3&3&          0.825803 E--11 -- &          0.142086 E--11 \\
0&2&2&2&         -- 0.958961 E--11 -- &          0.585948 E--12 \\
0&2&2&3&         -- 0.209232 E--10  + &          0.289031 E--11 \\
0&2&3&3&         -- 0.802359 E--11  + &          0.994701 E--12 \\
0&3&3&3&         -- 0.102576 E--09  + &          0.378476 E--10 \\
1&1&1&1&         -- 0.246426 E--10  + &          0.276326 E--10 \\
1&1&1&2&          0.915670 E--12 -- &          0.660629 E--12 \\
1&1&1&3&          0.303529 E--11 -- &          0.287480 E--11 \\
1&1&2&2&         -- 0.822697 E--11  + &          0.919635 E--11 \\
1&1&2&3&         -- 0.116294 E--11  + &          0.100024 E--11 \\
1&1&3&3&         -- 0.146918 E--10  + &          0.183799 E--10 \\
1&2&2&2&          0.908296 E--12 -- &          0.654735 E--12 \\
1&2&2&3&          0.109510 E--11 -- &          0.100875 E--11 \\
1&2&3&3&          0.717342 E--12 -- &          0.557293 E--12 \\
1&3&3&3&          0.450661 E--11 -- &          0.485065 E--11 \\
2&2&2&2&         -- 0.245154 E--10  + &          0.274313 E--10 \\
2&2&2&3&         -- 0.318500 E--11  + &          0.279750 E--11 \\
2&2&3&3&         -- 0.146317 E--10  + &          0.182912 E--10 \\
2&3&3&3&         -- 0.477335 E--11  + &          0.477368 E--11 \\
3&3&3&3&         -- 0.730168 E--10  + &          0.112865 E--09 \\
 \hline
\end{tabular}
\caption[Tensor components  of a massive rank $R=4$ six-point function]
{\label{tab:hexagonResults4rank}
Tensor components for a massive rank $R=4$ six-point function; kinematics
defined in Table \ref{phasespacepoint} and Figure \ref{fig:6pt5ptfig}.}
\end{table}

Selected tensor coefficients of five-point tensors for the case of
massive internal particles  are shown in Table \ref{5-point-massive}.\footnote{
Please notice that we show here five-point tensor \emph{coefficients},
while in the case of six-point tensors we have shown tensor
\emph{components}. The tensor components are representation independent
and should be preferred as numerical output. For the five-point
tensors with massive internal particles, however, we have arranged
for a one-to-one correspondence with output of \texttt{LoopTools 2.2},
so it might be interesting to have, in this case, the tensor coefficients
instead.}
The coefficients  have been compared with \texttt{LoopTools 2.2} and
indeed we agree. For the massive six-point functions, there
is no alternative package publicly available.

\begin{table}[tb]
\centering
\begin{tabular}{|l| r@{ i }l|}
\hline
$E_0$ &     0.702503 E-14  + &  0.170006 E-14   \\
\hline
$E_{1}$&    3.56379 E-15 -- & 5.58904 E-16 \\
\hline
$E_{12}$&-- 7.86411 E-16 +  & 1.03994 E-15 \\
\hline
$E_{00}$&-- 8.18587 E-11 +  & 1.80354 E-11\\
\hline
$E_{123}$&  3.51267 E-16 +  & 9.64413 E-17\\
\hline
$E_{001}$&  9.38702 E-12 +  & 2.18811 E-11 \\
\hline
\end{tabular}
\caption[]{\label{5-point-massive}
Selected tensor coefficients  of five-point tensor functions with massive
internal particles; kinematics defined in Table \ref{phasespacepoint}.
}
\end{table}

In presence of massless internal particles, we face potential infrared
singularities. Then, the loop functions are Laurent series in $\eps$,
starting with a term proportional to $\frac{1}{\epsilon^2}$, and one
has to care about re-normalizations compared to our basic definition
\ref{definition}.
A popular measure is \cite{Ellis:2007qk,Binoth:2008uq}: 
\begin{eqnarray} \label{norm2}
{\cal M} &=&
(\mu)^{4-d}~\frac{\Gamma(1 - 2\eps)}{\Gamma(1 + \eps) \Gamma^2(1 - \eps)}
 ~\int \frac{d^d k} {i \pi^{d/2}}.
\end{eqnarray}
When discussing Feynman integrals with a dependence on inverse powers
of $\epsilon$ there appears a dependence of their constant terms on
these conventions. For convenience of the reader, the tables are
produced with a normalization as introduced in Equation \ref{norm2},
with the choice $\mu=1$.

For the case of six-point and five-point functions with only massless
internal particles, we show only a few sample coefficients in Table
\ref{last1a} and Table \ref{last2a}, which are produced with our
Fortran package. The phase space point chosen here is defined in
Table \ref{massless-phasespacepoint}.  We checked that, within double
precision, we completely agree with corresponding numbers produced with
\texttt{golem95}.

Finally, to complete the list of relevant results, we show also sample
tensor coefficients for the case of both massive and massless internal
particles, for five-point tensors in Table \ref{5-point-mixed} and for
six-point tensors in Table \ref{6-point-mixed}.  For this case with
mixed internal masses, there is no other publicly released code available.

\bigskip

To summarize, we have presented in this article tensor integrals of
rank $R \leq 3$ for five-point functions and of  rank $R \leq 4$ for
six-point functions.  This is sufficient for the calculation of e.g. four
fermion production at the LHC with NLO QCD corrections.

There are further reactions of interest which will need higher-point
functions and higher  ranks of   five- and six-point functions.
The details of their reductions have been left for a later investigation.

\begin{table}[htb]
\centering
\begin{tabular}{|l| r@{  i }l | r@{  i }l | r@{.}l |}
\hline
&\multicolumn{2}{|c|}{$\epsilon^0$}
&\multicolumn{2}{|c|}{$1/\epsilon$}
&\multicolumn{2}{|c|}{$1/\epsilon^2$}\\
\hline
$F_0$    &-- 57.8724994 -- & 9248.84583 &-- 3167.69411 -- & 2981.57728 &-- 1003&89197    \\
\hline
$F_{3}$  &-- 867.761166 + &859.212722&  273.495904 + & 483.076108 & 153&767901\\
\hline
$F_{22}$ &83.1234074 -- &271.20343&-- 75.7263181 -- & 95.1508846 &-- 30&2874673\\
\hline
\hline
$F^{000}$ &-- 185.635891  + &  1465.754753 & 487.259427 + &    525.6914058 &174&2745041\\
\hline
$F^{1111}$ &-- 2.64116950 -- & 4.28827971 & -- 0.8480346995 -- &   0.4557274228 &-- 0&1450625441\\
\hline
\end{tabular}
\caption[]{\label{last2a}
Tensor coefficients $F_0, F_{3}, F_{22}$ and tensor components $F^{000},
F^{1111}$ of six-point functions; all internal particles are massless,
kinematics of Table \ref{massless-phasespacepoint}.
}
\end{table}

\begin{sidewaystable}
\centering
\begin{tabular}{|l| r@{ i }l | r@{ i }l | r@{ + i }l |}
\hline
&\multicolumn{2}{|c|}{$\epsilon^0$}
&\multicolumn{2}{|c|}{$1/\epsilon$}
&\multicolumn{2}{|c|}{$1/\epsilon^2$}\\
\hline
$E_0$     & 202.168496 + &3211.04072 &1022.10601 + &972.027061 &309.405823&0.0\\
\hline
$E_{2}$   & 264.996441 -- &303.068452  & -- 96.4696846 -- &149.228472   & -- 47.5008979&0.0\\
\hline
$E_{33}$ &1780.58042  + &  2914.50734&  927.71650 +  &   568.572069   &         180.982111 &0.0 \\
\hline
$E_{00}$ &9.56327810  + & 1.61648472E-13  &4.70734562E-14  + &  2.48689958E-14  &         7.10542736E-15&0.0 \\
\hline
$E_{444}$ & -- 1035.29689 -- &1422.01085& -- 452.640112 -- &  254.226520& -- 80.9228146 & 0.0 \\
\hline
$E_{001}$ & -- 0.81227772 -- & 5.68434189 E--14 & -- 2.04281037 E--14 -- & 2.84217094 E--14 & -- 7.10542736 E--15 &0.0\\
\hline
\end{tabular}
\caption[]{\label{last1a}
Selected tensor coefficients of five-point tensor functions
with massless internal particles; kinematics defined in Table
\ref{massless-phasespacepoint}.
}

\vspace*{5mm}

\centering
\begin{tabular}{|l| r@{ i }l | r@{ i }l | r@{ + i }l |}
\hline
&\multicolumn{2}{|c|}{$\epsilon^0$}
&\multicolumn{2}{|c|}{$1/\epsilon$}
&\multicolumn{2}{|c|}{$1/\epsilon^2$}\\
\hline
$E_0$  & -- 0.289852933 E+04 + &  0.228935552 E+03 & -- 0.945038648 E+02 + &   0.454178453 E+02&0.7112330546 E+01  &  0.0\\
\hline
$E^{3}$ &0.168344624 E+03 -- &   0.181758172 E+02 & 0.4242553725 E+01 -- &   0.338838829 E+01& -- 0.6442770877 E+00 &0.0\\
\hline
$E^{23}$&-- 0.79409571852 E+01  +  &  0.5445326927 E+00 & -- 0.3008645503 E+00  + &    0.9457613783 E--01&0.1027869989 E--01 &0.0 \\
\hline
$E^{012}$&0.2472148936 E+01 -- &  0.127011969 E+00 & 0.9699262574 E-01  -- &  0.2560545796 E--01& -- 0.2331885086 E--02&0.0 \\
\hline
$E^{2130}$& 0.2733228280 E+02  --  & 0.519106421 E+02 & -- 0.909476582 E+01   +   & 0.1744459753 E--02 & 0.2112313083 E--03&0.0 \\
\hline
\end{tabular}
\caption[]{\label{5-point-mixed}
Selected tensor components of five-point tensor functions with both massive
and massless internal particles; kinematics defined in
Table \ref{phasespacepoint-1}.}

\vspace*{5mm}

\centering
\begin{tabular}{|l| r@{ i }l | r@{ i }l | r@{ + i }l |}
\hline
&\multicolumn{2}{|c|}{$\epsilon^0$}
&\multicolumn{2}{|c|}{$1/\epsilon$}
&\multicolumn{2}{|c|}{$1/\epsilon^2$}\\
\hline
$F_0$  & 0.2403558675 E+04 -- &  0.2058213187 E+03 & 0.7315208677 E+02 --  &0.4276718518 E+02  &-- 0.7543148872 E+01& 0.0\\
\hline
$F^{2}$ &0.1112747404 E+03 -- & 0.6809282900 E+01   & 0.4419243474 E+01 --  & 0.1201033663 E+01&-- 0.1044856909 E+00&0.0\\
\hline
$F^{13}$&-- 0.1014018623 E+02 + &  0.1797332619 E+01  & -- 0.5914958485 E--01  +  &0.3275539398 E+00 &0.7678550480 E--01&0.0 \\
\hline
$F^{123}$&-- 0.5007216712 E+00  + &   0.4194342396 E--01&   -- 0.1642316924 E--01   +   &0.7789453935 E--02&0.1225024390 E--02&0.0\\
\hline
$F^{3210}$& 0.1263455978 E+00 -- &  0.6509987460 E--02 & 0.4610567958 E--02  -- & 0.1506637282 E--02& -- 0.1945123881 E--03& 0.0\\
\hline
\end{tabular}
\caption[]{\label{6-point-mixed}
Selected tensor components of six-point tensor functions with both massive
and massless internal particles; kinematics defined in
Table \ref{phasespacepoint-1}.}

\end{sidewaystable}

\clearpage

\section*{Acknowledgements}
Work supported by Sonderforschungsbereich/Transregio SFB/TRR 9 of DFG ``Com\-pu\-ter\-ge\-st\"utz\-te Theoretische Teilchenphysik"
and by the European Community's Marie-Curie Research Trai\-ning Networks
MRTN-CT-2006-035505 ``HEPTOOLS'' and MRTN-CT-2006-035482 ``FLA\-VIA\-net''.
K.K. acknowledges a scholarship from the UPGOW project co-financed by
the European Social Fund.
J.F. likes to thank DESY for the kind hospitality. 
We thank 
Th.~Binoth,
A.~Denner,
S.~Dittmaier,
Th.~Hahn,
C.~Papadopoulos
and
P.~Uwer
for useful discussions.

\clearpage

\appendix

\section{\label{signedminors}Gram determinants and algebra of signed minors}
In this section relations are derived, which will turn out to be
indispensable in our tensor reductions.

We begin with some notational remarks on Gram determinants $G_{n-1}$,
\begin{eqnarray}
 G_{n-1} &=& |2 q_jq_k|, ~~~j,k=1,\cdots, n-1.
\end{eqnarray}
The \emph{modified Cayley determinant} of a diagram
with $n$ internal lines  with chords $q_j$ is:
\begin{eqnarray}\label{gram1}
()_n &=&
\left|C_{jk}\right|, ~~~j,k=0,\cdots, n,
\\ \nonumber
&=&
\left|
\begin{array}{ccccc}
  0 & 1       & 1       &\ldots & 1      \\
  1 & Y_{11}  & Y_{12}  &\ldots & Y_{1n} \\
  1 & Y_{12}  & Y_{22}  &\ldots & Y_{2n} \\
  \vdots  & \vdots  & \vdots  &\ddots & \vdots \\
  1 & Y_{1n}  & Y_{2n}  &\ldots & Y_{nn}
\end{array}
\right|,
\label{MCD}
\end{eqnarray}
with
\bea
\label{gram}
Y_{jk}=-(q_j-q_k)^2+m_j^2+m_k^2.
\eea
From our choice $q_n=0$, it follows that both determinants are related:
\bea
()_n=-G_{n-1} ,
\eea
and we will usually call $()_n$ the Gram determinant of the Feynman integral.

\emph{Signed minors} \cite{Melrose:1965kb}
are determinants (with a sign convention) which are obtained by excluding rows and columns from the   modified Cayley determinant $()_n$.
They are denoted by the symbol
\bea\label{gram2}
\left(
\begin{array}{ccc}
  j_1 & j_2 & \cdots j_m\\
  k_1 & k_2 & \cdots k_m\\
\end{array}
\right)_n,
\eea
labelling the rows $j_1,j_2,\cdots, j_m$ and columns $k_1,k_2,\cdots, k_m$ which have been
excluded from $()_n$.
The sign of a signed minor is defined by
\begin{eqnarray}
(-1)^{j_1+j_2+ \cdots +j_m+k_1+k_2+ \cdots +k_m}\times \mathtt{Signature}[j_1,j_2, \cdots j_m]\times
\mathtt{Signature}[k_1,k_2, \cdots k_m] ,
\end{eqnarray}
where $\mathtt{Signature}$ gives the sign of permutations to place the
indices in increasing order.
This agrees e.g. with the definition of the operator $\mathtt{Signature[List]}$ in Mathematica.
As an example may serve the quantity $\Delta_n$:
\bea\label{gram3}
\Delta_n=  \left|
\begin{array}{cccc}
  Y_{11}  & Y_{12}  &\ldots & Y_{1n} \\
  Y_{12}  & Y_{22}  &\ldots & Y_{2n} \\
  \vdots  & \vdots  &\ddots & \vdots \\
  Y_{1n}  & Y_{2n}  &\ldots & Y_{nn}
\end{array}
         \right| = {\begin{pmatrix} 0 \\ 0 \end{pmatrix} }_n.
\label{mcd}
\eea

We now will derive two relations between signed minors.
Let us introduce
\bea 
A_{ij}^s \equiv
- {\begin{pmatrix} 0 \\ j \end{pmatrix} }_5 
  {\begin{pmatrix} 0&  s \\ 0&i \end{pmatrix}}_5 
  {\begin{pmatrix} s \\ s \end{pmatrix} }_5 
-
{\begin{pmatrix} s     \\ j \end{pmatrix} }_5 
{\begin{pmatrix} 0&  s \\ i&s \end{pmatrix} }_5 
{\begin{pmatrix} 0     \\ 0 \end{pmatrix} }_5 
+ 
{\begin{pmatrix} 0     \\ s \end{pmatrix} }_5  
{\begin{pmatrix} 0&  s \\ 0&s \end{pmatrix} }_5 
{\begin{pmatrix} i \\ j \end{pmatrix} }_5 \, .
\label{cancel4}
\eea
We are going to show that this expression can be factorized as
\bea
A_{ij}^s & = & {\begin{pmatrix} ~\\ ~\end{pmatrix} }_5 X_{ij}^s\, ,
\eea
and provide an explicit expression for$ X_{ij}^s$.
To begin with, we show that $ A_{ij}^s$ is symmetric in the indices
$i$ and $j$ for fixed $s$.
Obviously the third term on the right hand side of (\ref{cancel4}) is
symmetric since we consider a symmetric determinant. 
The symmetry of the first two terms means
\bea
{\begin{pmatrix} s \\ s \end{pmatrix} }_5 
\left[
{\begin{pmatrix} 0 \\ i \end{pmatrix} }_5 
{\begin{pmatrix}0&j \\  0&s \end{pmatrix} }_5
-
{\begin{pmatrix} 0 \\ j \end{pmatrix} }_5 
{\begin{pmatrix} 0&i \\ 0&s \end{pmatrix} }_5 
\right] +
{\begin{pmatrix} 0 \\ 0 \end{pmatrix} }_5 
\left[
{\begin{pmatrix} s \\ i \end{pmatrix} }_5 
{\begin{pmatrix} 0&  s \\ j&s\end{pmatrix} }_5
-
{\begin{pmatrix} s \\ j \end{pmatrix} }_5 
{\begin{pmatrix} 0&  s \\ i&s \end{pmatrix} }_5 \right]=0.
\label{vanish}
\nl
\eea
The first square bracket of (\ref{vanish}) can be evaluated using (A.13) of \cite{Melrose:1965kb}, i.e.
\bea
{\begin{pmatrix} 0 \\ j \end{pmatrix} }_5 {\begin{pmatrix} 0&i \\ 0&s \end{pmatrix} }_5=-{\begin{pmatrix} 0 \\ 0 \end{pmatrix} }_5 
{\begin{pmatrix} 0&  s \\ i&j \end{pmatrix}}_5+{\begin{pmatrix} 0 \\ i \end{pmatrix} }_5 {\begin{pmatrix}0&j \\  0&s \end{pmatrix} }_5
\eea
and (\ref{vanish}) then results in
\bea
{\begin{pmatrix} s \\ i \end{pmatrix} }_5 {\begin{pmatrix} 0&  s \\ j&s\end{pmatrix} }_5+{\begin{pmatrix} s \\ j \end{pmatrix} }_5 {\begin{pmatrix} 0&  s \\ s&i \end{pmatrix} }_5+{\begin{pmatrix} s \\ s \end{pmatrix} }_5 {\begin{pmatrix} 0&  s \\  i&j\end{pmatrix}}_5=0.
\label{circle}
\eea
This is proved by multiplication \footnote{Assuming here ${\left(  \right)_5} \ne 0$ means no limitation since
we are just looking for an algebraic relation. }
with ${\left(  \right)_5}$ and using Eqn. (A.8)  of \cite{Melrose:1965kb} with $r=2$, i.e.
\bea
{\begin{pmatrix} i&l\\j&k \end{pmatrix} }_5 {{\begin{pmatrix} ~\\ ~\end{pmatrix} }_5} = 
{\begin{pmatrix} i \\ j \end{pmatrix} }_5 
{\begin{pmatrix} l\\ k \end{pmatrix} }_5 - 
{\begin{pmatrix} i \\ k \end{pmatrix} }_5 
{\begin{pmatrix} l\\ j \end{pmatrix} }_5; ~~i,j,k,l = 0, \dots, 5.
\label{r2}
\eea
Inserting this, products of three factors of the form ${\begin{pmatrix} i \\ k \end{pmatrix} }_5$ cancel by pairs, q.e.d.~.

For the following, relations (A.11) and (A.12) of \cite{Melrose:1965kb} become important, i.e.
\bea
\sum_{i=1}^n {\begin{pmatrix} 0 \\ i \end{pmatrix} }_5 =()_5
\label{A11}
\eea
and
\bea
\sum_{i=1}^n {\begin{pmatrix} j \\ i \end{pmatrix} }_5 =0,~ (j \ne 0).
\label{A12}
\eea
Further, ``extensionals`` are needed, i.e. relations valid for $()_5$ can be extended to any minor
of $()_5$; an extensional of  (\ref{A11}) e.g. is
\bea
\sum_{i=1}^n 
{\begin{pmatrix} j&0 \\  k&i\end{pmatrix}}_5 
=
{\begin{pmatrix} j \\ k \end{pmatrix} }_5.
\label{A13}
\eea
As the simplest case we now immediately obtain from
(\ref{cancel4}) $A_{ss}^s=0$, i.e.
\bea
X_{ss}^s=0.
\label{simplest}
\eea
Applying  (\ref{A11}) and  (\ref{A12}) to  (\ref{cancel4}), we see
\bea
\sum_{j=1}^5 A_{ij}^s = -
{{\begin{pmatrix} ~\\ ~\end{pmatrix} }_5} 
{\begin{pmatrix} 0&  s \\ 0&i\end{pmatrix} }_5  
{\begin{pmatrix} s \\ s \end{pmatrix} }_5
\label{sumj}
\eea
and due to the symmetry in $i$ and $j$ we also have
\bea
\sum_{i=1}^5 A_{ij}^s = -{{\begin{pmatrix} ~\\ ~\end{pmatrix} }_5} 
{\begin{pmatrix} 0&  s \\ 0&j \end{pmatrix} }_5  
{\begin{pmatrix} s \\ s \end{pmatrix} }_5,
\label{sumi}
\eea
which gives us a hint of how $X_{ij}^s$ might look, namely due to
(\ref{sumj}) it should contain
a term $-{\begin{pmatrix} 0&  s \\0&i \end{pmatrix} }_5 
{\begin{pmatrix} 0&  s \\ j&s\end{pmatrix} }_5$.
A further contribution must vanish after summing over $i$.
Due to (\ref{simplest}) it must contain a factor 
$ 
{\begin{pmatrix}0&j \\ s&i \end{pmatrix} }_5
$ 
\footnote{Observe that
$
\sum_{j=1}^5   
{\begin{pmatrix}0&j \\ s&i \end{pmatrix} }_5=0
$ 
but 
$\sum_{i=1}^5   {\begin{pmatrix}0&j \\ s&i \end{pmatrix} }_5
=
-\sum_{i=1}^5  
{\begin{pmatrix}0&j \\ i&s \end{pmatrix} }_5
=
- 
{\begin{pmatrix} j \\ s \end{pmatrix} }_5
$. 
}. 
The second factor of this contribution can only be depend on $s$ and
has been determined by explicit calculation to be ${\begin{pmatrix} 0&  s \\ 0&s \end{pmatrix} }_5$. Thus we conclude:
\bea
X_{ji}^s=X_{ij}^s=
-
{\begin{pmatrix} 0&  s \\ 0&i \end{pmatrix}}_5 
{\begin{pmatrix} 0&  s \\ j&s\end{pmatrix} }_5
+ 
{\begin{pmatrix}0&j \\  s&i \end{pmatrix} }_5 
{\begin{pmatrix} 0&  s \\ 0&s \end{pmatrix} }_5.
\label{Xijs}
\eea

We come now to the second relation between signed minors.
While (\ref{Xijs}) will be needed for the reduction of $4$-point tensors to scalars $I_4^s$,
for the reduction of $3$-point tensors to scalars $I_3^{st}$ we also need
\bea
-
{\begin{pmatrix} 0 \\ j \end{pmatrix} }_5 
{\begin{pmatrix} t&  s \\ 0&i\end{pmatrix} }_5 
{\begin{pmatrix} s \\ s \end{pmatrix} }_5
-
{\begin{pmatrix} s \\ j \end{pmatrix} }_5 
{\begin{pmatrix} t&  s \\ i&s \end{pmatrix} }_5 
{\begin{pmatrix} 0 \\ 0 \end{pmatrix} }_5 
+ 
{\begin{pmatrix} 0 \\ s \end{pmatrix} }_5  
{\begin{pmatrix} t&  s \\ 0&s \end{pmatrix} }_5 
{\begin{pmatrix} i \\ j \end{pmatrix} }_5
={\begin{pmatrix} ~\\ ~\end{pmatrix} }_5 X_{ij}^{st}, \non \\
\label{cancel3}
\eea
where again we have to show that indeed ${\left(  \right)_5}$ factorizes and we have to give an
explicit expression for $X_{ij}^{st}$.
The left-most term on the left hand side is an auxiliary term.
It is antisymmetric in $s$ and $t$
after the cancellation of ${\begin{pmatrix} s \\ s \end{pmatrix} }_5$
and
vanishes after summation over $s$ and $t$ because $I_3^{st}$ is symmetric in $s$ and $t$.
The cancellation of ${\begin{pmatrix} s \\ s \end{pmatrix} }_5$ has to be checked explicitly in every case where
(\ref{cancel3}) is applied.

We observe that the expressions for $X_{ij}^s$ (\ref{cancel4}) and
$X_{ij}^{st}$ (\ref{cancel3}) differ only by replacing one $0$ by $t$. Therefore the following ansatz is implied for $X_{ij}^{st}$.
\bea\label{xijst}
X_{ij}^{st}=  -
{\begin{pmatrix} 0&  s \\ 0&j \end{pmatrix} }_5 
{\begin{pmatrix} t&  s \\ i&s \end{pmatrix} }_5
+ 
{\begin{pmatrix} 0&i \\s&j \end{pmatrix} }_5 
{\begin{pmatrix} t&  s \\ 0&s \end{pmatrix} }_5.
\label{Yijst}
\eea
Now we directly evaluate $X_{ij}^{st} ()_5$ using (\ref{r2}):
\bea
X_{ij}^{st} ()_5=-\left[{\begin{pmatrix} 0 \\ 0 \end{pmatrix} }_5 {\begin{pmatrix} s \\ j \end{pmatrix} }_5-{\begin{pmatrix} s \\ 0 \end{pmatrix} }_5
{\begin{pmatrix} 0 \\ j \end{pmatrix} }_5 \right] {\begin{pmatrix} t&  s \\ i&s \end{pmatrix} } + \left[{\begin{pmatrix} 0 \\ s \end{pmatrix} }_5
{\begin{pmatrix} i \\ j \end{pmatrix} }_5 - {\begin{pmatrix} i \\ s \end{pmatrix} }_5 {\begin{pmatrix} 0 \\ j \end{pmatrix} }_5 \right] {\begin{pmatrix} t&  s \\ 0&s \end{pmatrix} }
\nl
\eea
and the remaining equation to be verified is
\bea
{\begin{pmatrix} s \\ 0 \end{pmatrix} }_5 {\begin{pmatrix} 0 \\ j \end{pmatrix} }_5 {\begin{pmatrix} t&  s \\ i&s \end{pmatrix} } - {\begin{pmatrix} i \\ s \end{pmatrix} }_5
{\begin{pmatrix} 0 \\ j \end{pmatrix} }_5 {\begin{pmatrix} t&  s \\ 0&s \end{pmatrix} } = -{\begin{pmatrix} 0 \\ j \end{pmatrix} }_5 {\begin{pmatrix} t&  s \\0&i \end{pmatrix} }
{\begin{pmatrix} s \\ s \end{pmatrix} }_5,
\eea
which is done by multiplying again with $()_5$ and again using (\ref{r2}).
This gives us at the same time also a more general proof for $X_{ij}^s$ (\ref{Xijs}), putting $t=0$.

\section{\label{sub1}
Reduction of dimensionally shifted five- and four-point integrals}
In this appendix we provide explicitly the needed recursion relations
for the reduction
of the five- and four-point functions. 
In spite of the fact that here, essentially, only two different
relations of \cite{Fleischer:1999hq} are applied for different indices and dimension,
namely (I.30) and (I.31), we consider it helpful and sometimes even necessary, to provide them in
detail. 
A special case of (I.31) is (\ref{A401}).
The others are
special cases of (I.30). 
For the six-point function relation (I.26) plays a major role and will
be quoted when applied.
\begin{eqnarray}
\label{A533}
{\nu}_{ijk} I_{5,ijk}^{[d+]^3}&=&-\frac{
{\begin{pmatrix} 0 \\ k \end{pmatrix} }_5}{{\begin{pmatrix} ~\\ ~\end{pmatrix} }_5} I_{5,ij}^{[d+]^2} +
 \sum_{s=1,s \ne i,j}^{5} \frac{
{\begin{pmatrix} s \\ k \end{pmatrix} }_5}{{\begin{pmatrix} ~\\ ~\end{pmatrix} }_5} I_{4,ij}^{[d+]^2,s} +
 \frac{
{\begin{pmatrix} i \\ k \end{pmatrix} }_5}{{\begin{pmatrix} ~\\ ~\end{pmatrix} }_5} I_{5,j}^{[d+]^2}+
 \frac{
{\begin{pmatrix} j \\ k \end{pmatrix} }_5}{{\begin{pmatrix} ~\\ ~\end{pmatrix} }_5} I_{5,i}^{[d+]^2}, \\
\label{A522}
{\nu}_{ij} I_{5,ij}^{[d+]^2}&=&-\frac{
{\begin{pmatrix} 0 \\ j \end{pmatrix} }_5}{{\begin{pmatrix} ~\\ ~\end{pmatrix} }_5} I_{5,i}^{[d+]} +
 \sum_{s=1,s \ne i}^{5} \frac{
{\begin{pmatrix} s \\ j \end{pmatrix} }_5}{{\begin{pmatrix} ~\\ ~\end{pmatrix} }_5} I_{4,i}^{[d+],s} +
 \frac{
{\begin{pmatrix} i \\ j \end{pmatrix} }_5}{{\begin{pmatrix} ~\\ ~\end{pmatrix} }_5} I_{5}^{[d+]}
.
\end{eqnarray}

The four-point function's shift is (I.44):
\begin{eqnarray}
\label{A411}
I_{4,i}^{[d+],s}&=&-
\frac{
{\begin{pmatrix} 0&  s \\ i&s \end{pmatrix} }_5
}
{
{\begin{pmatrix} s \\ s \end{pmatrix} }_5} I_{4}^{s} +
 \sum_{t=1,t \ne s}^{5} 
\frac{
{\begin{pmatrix} t&  s \\ i&s \end{pmatrix} }_5
}{
{\begin{pmatrix} s \\ s \end{pmatrix} }_5} I_{3}^{st},
\end{eqnarray}
and the four-point integrals occurring in the reduction are (I.50):
\begin{eqnarray}
\label{A401}
I_{4}^{[d+],s}&=&\left[\frac{{\begin{pmatrix} 0&  s \\ 0&s \end{pmatrix} }_5}{{\begin{pmatrix} s \\ s \end{pmatrix} }_5} I_{4}^{s}-
 \sum_{t=1,t \ne s}^{5} \frac{{\begin{pmatrix} t&  s \\ 0&s \end{pmatrix} }_5}{{\begin{pmatrix} s \\ s \end{pmatrix} }_5} I_{3}^{st}\right]\frac{1}{d-3} 
\end{eqnarray}
In applications we can put $d=4$ since $I_4^{[d+]}$ is UV- and IR-finite. 
Beyond that,
as it is done frequently \cite{Binoth:2005ff}, $I_4^{[d+]}$ can be used as well as a ``master integral''
(see e.g. (\ref{E00a})) without reduction to the generic dimension.

\clearpage

\bibliographystyle{JHEP}
\addcontentsline{toc}{section}{References}
\bibliography{2loops_teo}
\end{document}